\documentclass[pdflatex,sn-mathphys-num]{sn-jnl}

\usepackage{graphicx}%
\usepackage{multirow}%
\usepackage{amsmath,amssymb,amsfonts}%
\usepackage{amsthm}%
\usepackage{mathrsfs}%
\usepackage[title]{appendix}%
\usepackage{xcolor}%
\usepackage{textcomp}%
\usepackage{manyfoot}%
\usepackage{booktabs}%
\usepackage{algorithm}%
\usepackage{algorithmicx}%
\usepackage{algpseudocode}%
\usepackage{listings}%
\theoremstyle{thmstyleone}%
\theoremstyle{thmstyletwo}%

\theoremstyle{thmstylethree}%

\begin{document}

\title{Entropy-based measure of rock sample heterogeneity derived from micro-CT images}


\author*[1]{\fnm{Luan} \sur{Coelho Vieira da Silva}}\email{luan@cos.ufrj.br}
\author[1]{\fnm{Júlio} \sur{de Castro Vargas Fernandes}} 
\author[1]{\fnm{Felipe} \sur{Bevilaqua Foldes Guimarães}}
\author[1]{\fnm{Pedro} \sur{Henrique Braga Lisboa}}
\author[1]{\fnm{Carlos Eduardo} \sur{Menezes dos Anjos}}
\author[2]{\fnm{Thais} \sur{Fernandes de Matos}}
\author[2]{\fnm{Marcelo} \sur{Ramalho Albuquerque}}
\author[2]{\fnm{Rodrigo} \sur{Surmas}}
\author[1]{\fnm{Alexandre} \sur{Gonçalves Evsukoff}}

\affil[1]{COPPE - Federal University of Rio de Janeiro, mailbox 68506, Rio de Janeiro, 21941-972, Rio de Janeiro, Brazil}
\affil[2]{CENPES - Petrobras, Av. Horácio Macedo, 950, Rio de Janeiro, 21941-915, Rio de Janeiro, Brazil}


\abstract{
This study presents an automated method for objectively measuring rock heterogeneity via raw X-ray micro-computed tomography (micro-CT) images, thereby addressing the limitations of traditional methods, which are time-consuming, costly, and subjective. Unlike approaches that rely on image segmentation, the proposed method processes micro-CT images directly, identifying textural heterogeneity. The image is partitioned into subvolumes, where attributes are calculated for each one, with entropy serving as a measure of uncertainty. This method adapts to varying sample characteristics and enables meaningful comparisons across distinct sets of samples. It was applied to a dataset consisting of 4,935 images of cylindrical plug samples derived from Brazilian reservoirs. The results showed that the selected attributes play a key role in producing desirable outcomes, such as strong correlations with structural heterogeneity. To assess the effectiveness of our method, we used evaluations provided by four experts who classified 175 samples as either heterogeneous or homogeneous, where each expert assessed a different number of samples. One of the presented attributes demonstrated a statistically significant difference between the homogeneous and heterogeneous samples labelled by all the experts, whereas the other two attributes yielded nonsignificant differences for three out of the four experts. The method was shown to better align with the expert choices than traditional textural attributes known for extracting heterogeneous properties from images. This textural heterogeneity measure provides an additional parameter that can assist in rock characterization, and the automated approach ensures easy reproduction and high cost-effectiveness.

\section*{Article highlights}
\begin{itemize}
   \item An automated method for measuring rock textural heterogeneity using only micro-CT images without human evaluations.
    \item Lithological variations, diagenetic minerals, porous structures, and artifacts have the greatest influence on the measure.
   \item Our method outperforms traditional texture attributes in capturing distinct behaviors of both heterogeneous and homogeneous groups.

\end{itemize}
}


\maketitle

\section{Introduction}\label{sec_introduction}


Reservoir characterization is crucial to understanding the petrophysical properties and determining the viability of exploration and production activities, such as identifying optimal drilling locations and designing enhanced oil recovery methods. Reservoir heterogeneity can then be defined according to the variations exhibited by these properties as a function of space \citep{ahmed2010fundamentals}. The traditional heterogeneity assessment methods include visually inspecting core samples and using coefficients that are statistically derived from well logs. Techniques such as the coefficient of variation, the Lorenz coefficient \citep{fitch2015integrated}, and the Dykstra-Parsons coefficient \citep{dykstra1950prediction} summarize heterogeneity with single-value coefficients. However, these methods overlook the detailed spatial distributions of rock components and petrophysical property discontinuities \citep{sahu2024assessment}, leading to an incomplete understanding of formation heterogeneity.

Determining petrophysical properties in a laboratory can be time-consuming, costly, and potentially destructive \citep{zamanzadeh2024robust}. The patterns exhibited by the variability of rock textures and geological features change depending on the chosen scale. Different forms of heterogeneity may be present at each measurement level. High-resolution imaging technologies have enabled the detailed capture of rock textures and compositions at various scales, leading to a substantial number of digital rock physics studies aimed at predicting petrophysical properties and rock characterizations being published in recent years \citep{dos2023permeability, dos2021deep, de2024absolute, liu2023hierarchical}. However, challenges remain, particularly concerning the trade-off between the image resolution and field of view. When the imaging resolution exceeds the smallest pore size, subresolution porosity can significantly impact the obtained petrophysical property estimates (including the assessment of heterogeneity), as fine-scale variations may be overlooked \citep{carrillo2022impact}.

The proposed methodology involves systematically dividing the input images into subvolumes and calculating statistical attributes, such as the mean and standard deviation, within each subvolume. Shannon's entropy \citep{shannon1948mathematical} is then applied to these attributes to quantify their variability and assess their degree of heterogeneity, with higher entropy values indicating greater unpredictability or heterogeneity. Finally, quantiles are applied to the entropy values to provide a more interpretable measure of the heterogeneity observed across the dataset.

This approach is designed to address the need for an automated, objective method that can directly quantify rock heterogeneity from the texture of micro-computed tomography (micro-CT) images, bypassing the need for segmentation. The traditional methods for analysing rock heterogeneity often rely on laboratory measurements or expert interpretation, both of which can be time-consuming, subjective, and prone to variability. Moreover, segmentation approaches, which are commonly used in other methods, depend heavily on the input image resolution and the employed segmentation technique, introducing other potential sources of error. Our method overcomes these challenges by working directly with micro-CT images and using a systematic approach to quantify heterogeneity. By doing so, petrophysical experts are provided with an easy and rapid additional metric, i.e., a measure of texture heterogeneity that can support their analyses by enabling comparisons across values derived from different images. However, as a measure of textural heterogeneity, this metric does not always align with the geological concept of sample heterogeneity.

The remainder of this paper is organized as follows. Section \ref{sec_related_works} reviews the related works. Section \ref{sec:materials} outlines the materials, methods, and metrics used. Section \ref{sec:results} presents the results. Section \ref{sec:conclusion} concludes with a discussion and suggestions for future work.

\section{Related works}\label{sec_related_works}

One rock heterogeneity characterization approach involves identifying related features. \citet{jones1987using} proposed the notion that an inertial coefficient substantially greater than the characteristic length indicates a sample that is nonhomogeneous with respect to permeability. \citet{li2004characterization} employed the fractal dimension to quantify heterogeneity, as it represents a measure of the complexity and irregularities exhibited by rock pore structures. A positive correlation with heterogeneity was qualitatively determined by comparing pore size distributions. However, both of these approaches rely on laboratory measurements (since the fractal dimension is determined from capillary pressure curves), which can be costly.

\citet{lin2021prediction} introduced a manually labelled homogeneous coefficient for quantifying rock heterogeneity on the basis of CT scans. Building on this research, \citet{tian2023quantification} used image recognition to predict rock properties, incorporating the previously proposed coefficient. Their methodology involves a convolutional neural network (CNN) that includes a 3D spatial grid based on the observed mineral distribution. Detailed information concerning the spatial arrangement of rock components is provided, allowing the CNN to predict coefficients for new samples. By incorporating the rock homogeneity coefficient, the loss values induced when predicting rock strength and elastic modulus values were reduced by 25.7\% and 3.9\%, respectively. However, the training data employed for the homogeneity coefficient consisted of prelabelled samples with coefficients discretized into only six values, adding subjectivity.

\citet{mohamed2023scale} extended an existing method to automate the heterogeneity estimation process at the pore scale for two-phase media; this approach is independent of the input image resolution. The algorithm calculates porosity within a moving window at various radii based on the maximum solid-to-pore distance observed in the given 3D image. Porosity variances are then plotted against the radius index, providing a heterogeneity measure that is influenced by these radius-based bounds, with larger radii generally leading to smaller variations. While the method can incorporate other features, the presented results focused on the spatial variability of porosity to obtain variability measures. \citet{yazynina2021new} categorized rock heterogeneity based on pore size and pore arrangement variability. By setting a distinct threshold for each parameter, they divided the observed heterogeneity into four distinct areas. Both of the above approaches, despite being designed to quantify heterogeneity at the pore scale, require segmentation to implemented beforehand, which introduces a new source of error due to the limited accuracy of this process \citep{xie2020sesv}.

\citet{altieri2019advances} identified two approaches for estimating spatial entropy. The first method accounts for unequally partitioned subareas and is computed only for binary variables, indicating the presence or absence of an attribute at each location. The second approach introduces spatial considerations by transforming the examined variable to account for the distances between realizations (cooccurrences). One area that extensively applies spatial entropy involves the quantification and characterization of landscape patterns, where entropy-related metrics have rapidly developed into efficient tools for this purpose \citep{8963966, vranken2015review, wang2018spatial}.

Given the advancements achieved in terms of high-resolution imaging technologies, texture analysis has emerged as a promising option for assessing heterogeneity via automated approaches. It describes the spatial arrangement and relationships of the components of a rock, thereby aiding in identifying variability and heterogeneity. Recently, \citet{sahu2024assessment} proposed a methodology that uses a grey-level cooccurrence matrix (GLCM) to extract textural attributes and captures additional compositional attributes from well logs. Principal component analysis \citep{abdi2010principal} is applied to these attributes to represent the variability at each depth with a unique value. Finally, an index is analytically calculated to quantify the heterogeneity at each depth. While this method is applied on a macro scale, it underscores the potential of texture analysis for assessing rock heterogeneity, since applying this approach at a finer scale with more details could yield even better results.

\section{Materials and methods}\label{sec:materials}

\subsection*{Dataset}\label{subsuc:dataset}

The study was carried out using 4,744 cylindrical plug samples collected from the oilfields of Brazil. Most of the samples are carbonates, but sandstone samples are also present. The dataset consisted of 4,935 micro-CT images, as some samples had multiple images acquired at different resolutions and/or under varying rock preparation conditions. The image resolution ranged from 13.00 \textmu m to 64.59 \textmu m. Micro-CT images were generated directly through a reconstruction process conducted via the VTomex L300 system manufactured by Baker Hughes. The images underwent a subjective quality control process, where factors such as noise levels and the presence of artefacts, including beam hardening, were evaluated. The median energy and source current used for the scanning process were 160 kiloelectronvolts and 40 microamperes, respectively.

\subsection*{Proposed approach}

The proposed method for quantifying rock sample heterogeneity is summarized by the \texttt{RankSample} procedure in Algorithm \ref{alg:entropy_heterogeneity}. First, the \texttt{DivideImageIntoSubcubes} function splits each 3D micro-CT image into subcubes based on the specified number of divisions along the x and y coordinate axes. The \texttt{CalculateAttributes} function then computes the statistical attributes of each subcube, which are subsequently standardized. Following this, the \texttt{CalculateEntropy} function calculates the Shannon entropy of each image. For continuous attributes, kernel density estimation is applied to estimate the corresponding probability density functions, whereas for discrete attributes, empirical probability mass functions are used. The resulting entropy values produced for the set of images are stored in an entropy database. Finally, the \texttt{Rank} function uses this database to rank the images based on their entropy quantile probabilities. The output of this process is an entropy-based heterogeneity coefficient, which provides an automatic and objective rock sample textural heterogeneity measure determined using only micro-CT images.

\begin{algorithm}
\caption{Entropy-based assessment of rock sample heterogeneity}
\label{alg:entropy_heterogeneity}
\begin{algorithmic}[1]
\State \textbf{Input:} Set of 3D micro-CT images $I$ restricted to the cuboid VOI (dimensions: $dim_x \times dim_y \times dim_z$, with $dim_x = dim_y$), number of divisions along the x and y coordinates $d$, and attribute $f$.
\State \textbf{Output:} Entropy-based heterogeneity coefficient $H$

\Procedure{RankSample}{$I, d, f$}
    \State $subcubes \gets$ \textsc{DivideImageIntoSubcubes}($I, d$)
    \State $standardizedAttributes \gets$ \textsc{CalculateAttributes}($subcubes, f$)
    \State $entropyDatabase \gets$ \textsc{CalculateEntropy}($standardizedAttributes$)
    \State $H \gets$ \textsc{Rank}($entropyDatabase, d, f$)
    \State \Return $H$
\EndProcedure
\end{algorithmic}
\end{algorithm}

\begin{figure}[H]
    \centering
    \begin{tabular}{ccc}
        \includegraphics[width=0.3\textwidth]{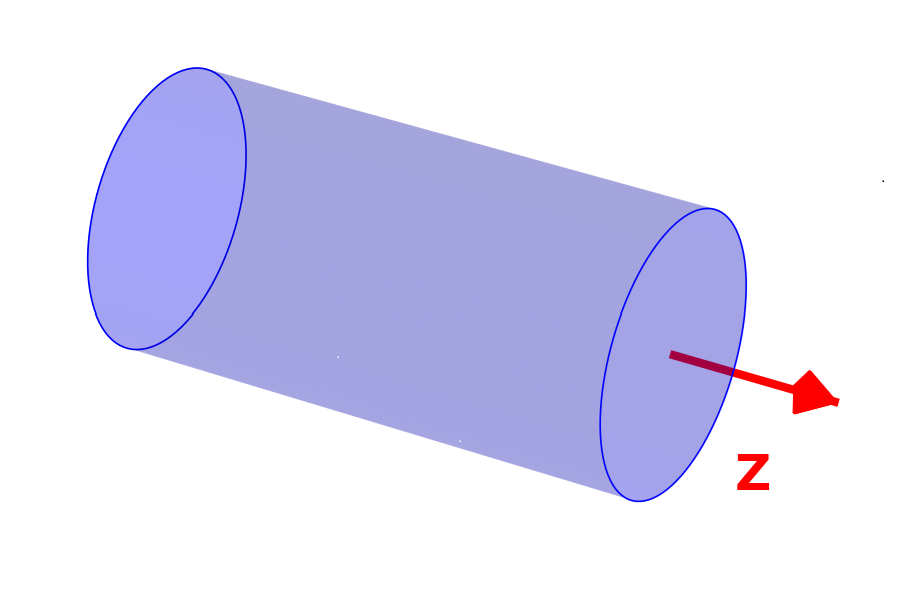} &
        \includegraphics[width=0.3\textwidth]{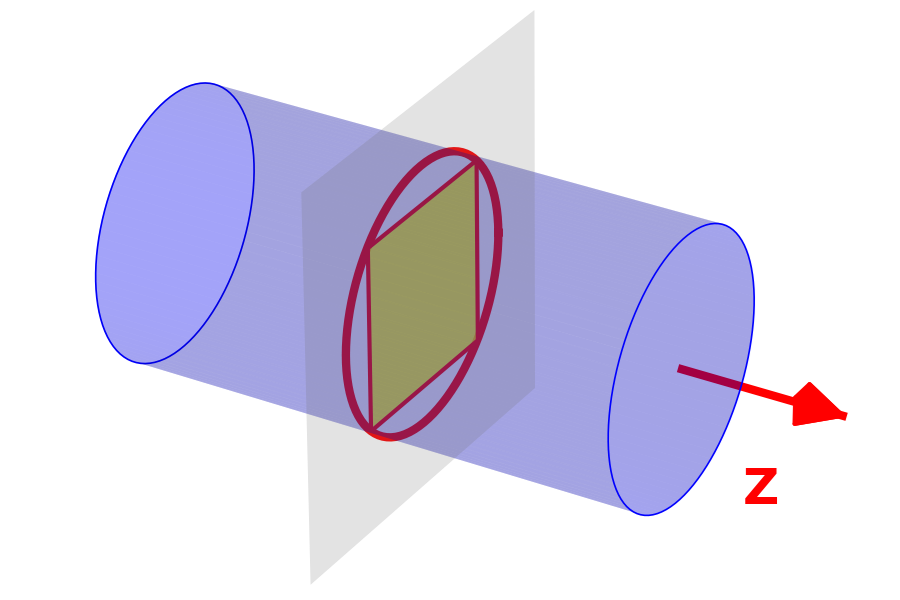} &
        \includegraphics[width=0.3\textwidth]{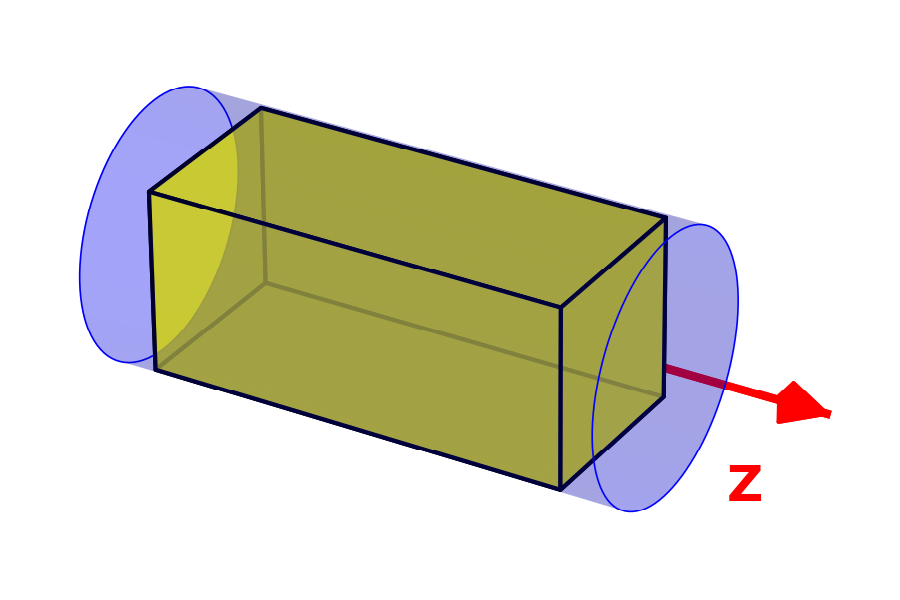} \\
        \parbox{0.3\textwidth}{\centering \small (a) The representation of the original cylindrical rock sample.} &
        \parbox{0.3\textwidth}{\centering \small (b) Cropping to the largest possible square area within the horizontal cross-section.} &
        \parbox{0.3\textwidth}{\centering \small (c) Expanding the VOI along the z-dimension to maximize the volume.} \\
    \end{tabular}
    \caption{Illustration of the steps used to define the VOI.}
    \label{fig:method_VOI}
\end{figure}

To apply this method, the empty space surrounding the rock sample in the input image must first be removed by defining a cuboid volume of interest (VOI), as illustrated in Figure \ref{fig:method_VOI}. The process begins with the representation of the cylindrical rock sample, as depicted in Figure \ref{fig:method_VOI}a. For each micro-CT image, the void area around the cylindrical sample is removed by cropping the horizontal cross-section to the largest possible square area that fits within the circular boundary, as shown in Figure \ref{fig:method_VOI}b. The z-dimension is then expanded to include the maximum possible volume of the rock within the cropped region while ensuring the exclusion of voids caused by misalignment, crookedness, or any external structures used to stabilize the sample during imaging (Figure \ref{fig:method_VOI}c).

\begin{figure}[H]
    \centering
    \begin{tabular}{cc}
        \includegraphics[width=0.4\textwidth]{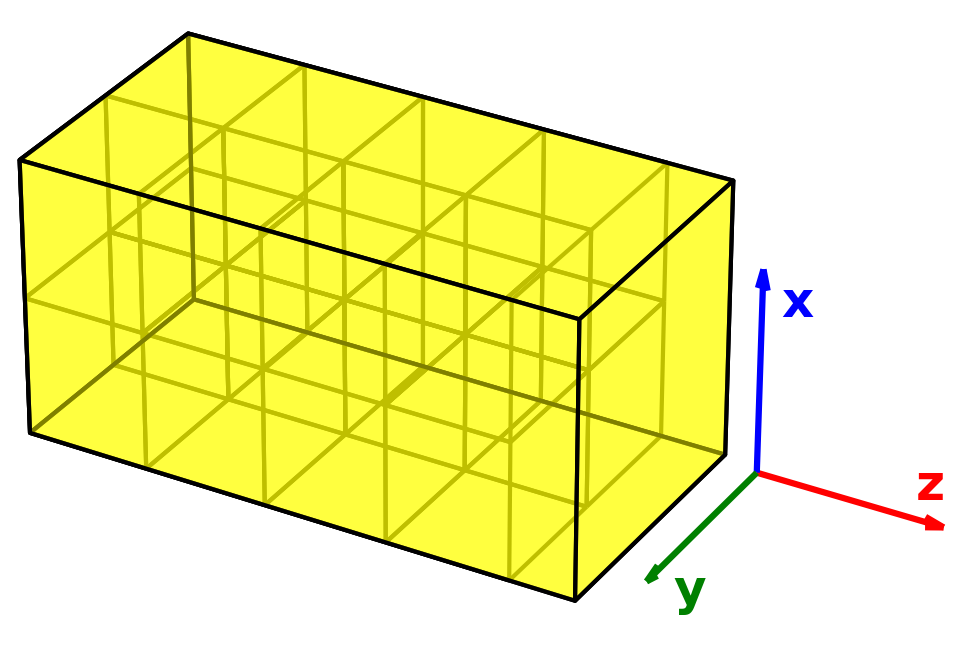} & 
        \includegraphics[width=0.4\textwidth]{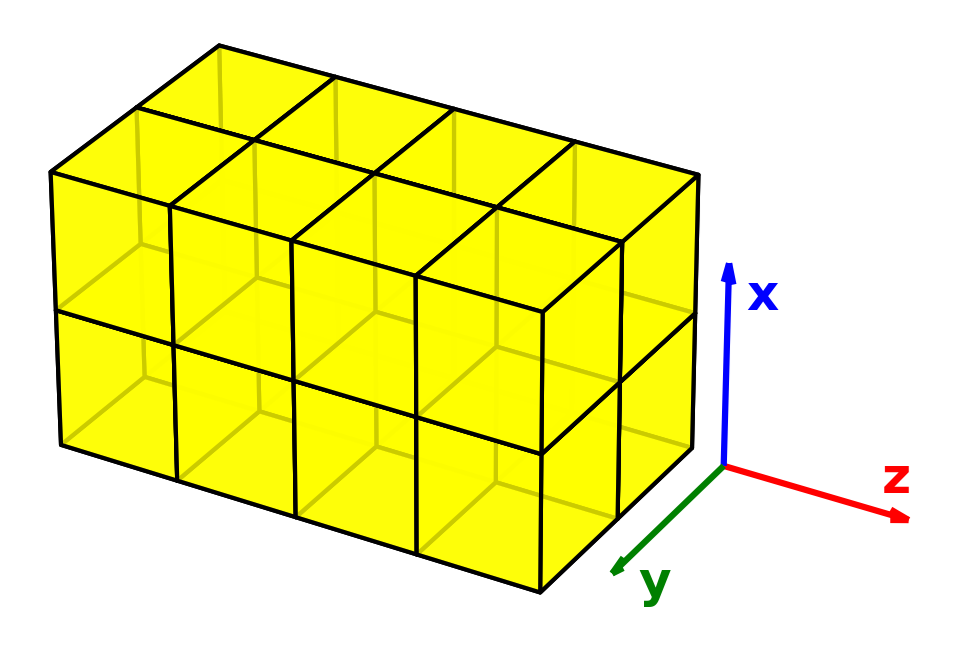} \\
        \parbox{0.4\textwidth}{\centering \small (a1) The VOI is divided into subcubes for 2 division segments along the x and y dimensions.} &
        \parbox{0.4\textwidth}{\centering \small (a2) Subcubes selected for the 2 division segments.} \\
        \includegraphics[width=0.4\textwidth]{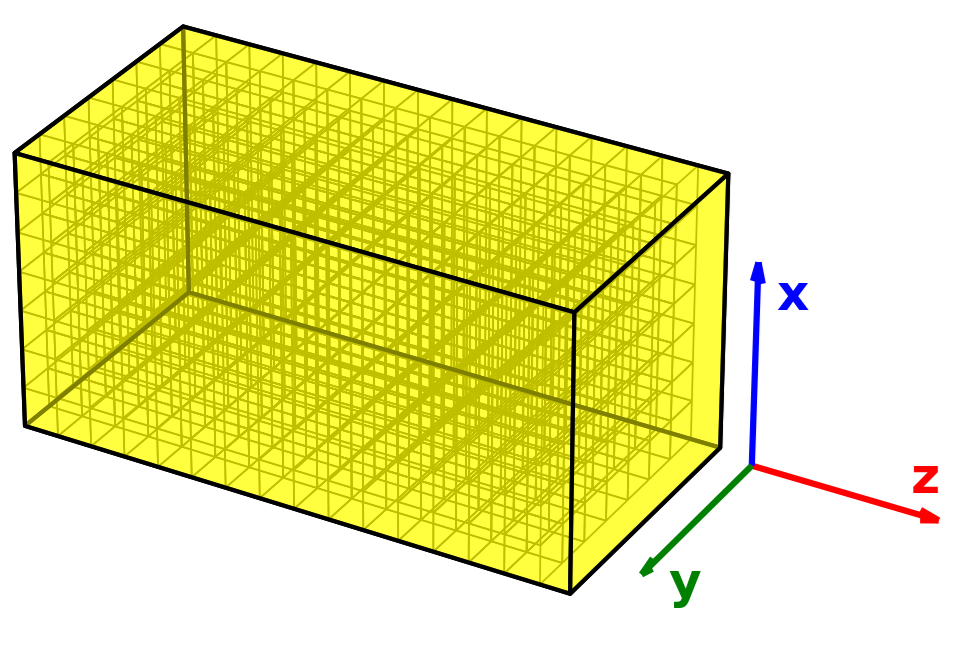} & 
        \includegraphics[width=0.4\textwidth]{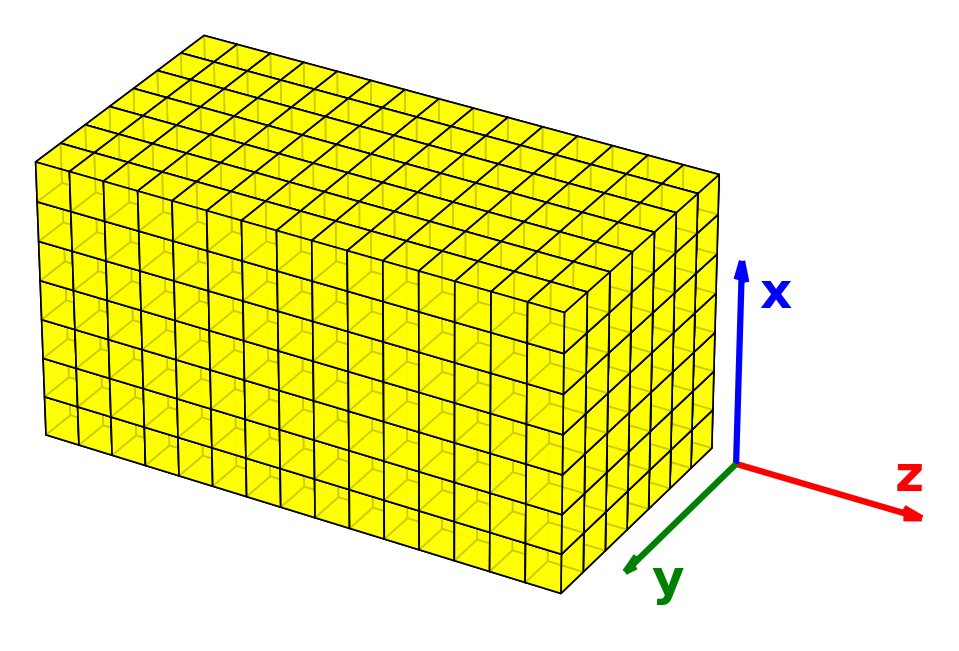} \\
        \parbox{0.4\textwidth}{\centering \small (b1) The VOI is divided into subcubes for 7 segments along the x and y dimensions.} &
        \parbox{0.4\textwidth}{\centering \small (b2) Subcubes selected for 7 segments.} \\
    \end{tabular}
    \caption{Illustration of the VOI division and subcube selection processes.}
    \label{fig:method_subcubes}
\end{figure}

The sample images, now restricted to the VOI, are subdivided into equal-sized cubic parts by performing division operations along the three spatial dimensions. The x- and y-dimensions are divided a specified number of times to determine the segment size. This segment size then dictates the number of divisions along the z-dimension, defining the total number of subcubes contained within the sample, as shown in Figure \ref{fig:method_subcubes}. A portion of the VOI derived from the sample may remain unused, depending on how well the subcubes fit within the volume. Figures \ref{fig:method_subcubes}a1 and \ref{fig:method_subcubes}a2 illustrate that dividing the sample into 2 segments per axis may leave a considerable volume not covered by subcubes. A more refined division scheme with 7 segments per axis (Figures \ref{fig:method_subcubes}b1 and \ref{fig:method_subcubes}b2) provides a more detailed subdivision result. In the x- and y-directions, within each subcube, statistical attributes are calculated directly from the grayscale values.

Whether discrete or continuous, any attribute that can be applied to the grayscale values of the micro-CT images can be used; however, we focus on simple attributes that are straightforward and computationally efficient to derive (such as the mean, standard deviation, median, and maximum). Once an attribute is chosen, its corresponding value is calculated for all subcubes. These values are then standardized by considering all samples across a selected number of divisions and a chosen attribute. Standardization matters because although discrete Shannon entropy is invariant to scaling, this does not hold in the continuous case where kernel density estimation is applied. By standardizing the values, we ensure that the entropy calculations remain unaffected by issues related to varying scales, preserving the integrity of the analysis. In particular, we note that the entropy results calculated based on 8-bit and 16-bit images yield significantly different results if attribute standardization is not applied.

The concept of entropy, as defined by Shannon, measures the amount of uncertainty or information contained in a system \citep{ali2023shannon}. In the context of micro-CT rock images, entropy quantifies the variability or heterogeneity within a given set of subvolumes (subcubes) based on the selected attribute. Higher entropy values indicate greater variability, meaning that the subvolumes present more diverse and heterogeneous characteristics, whereas lower entropy values suggest greater uniformity.

In our approach, we calculate the entropy between the subcubes of a 3D image, with each subcube \(s \) assigned a value based on the selected attribute, such as the minimum, maximum, or mean. For discrete attributes, such as the minimum or maximum grayscale value, the possible values are limited to the grayscale range of the input image. We define \(a(s) \) as the value of the attribute $a$ applied to each subcube \(s \). Then, \(p(k) \) represents the empirical probability mass function of a value \(k \), where \(k \) is an element of the image of \(a(s) \), denoted as \( \text{Im}(a(s)) \), which is the set of all values that the function \(a \) assigns to the subcubes. The Shannon entropy \(H(I) \) is then measured in bits and is calculated by summing over all the possible values assigned to the subcubes across the image:

\begin{equation}
\begin{aligned}
H(I) = -\sum_{k \in \text{Im}(a(s))} p(k) \log_2(\text{p}(k))
\end{aligned}
\label{eq:shannon-discrete}
\end{equation}

For continuous attributes, such as the mean or standard deviation, it is not possible to limit their values to a discrete set. In this case, we estimate the probability density function \(f(k) \) via the kernel density estimation \citep{wkeglarczyk2018kernel} process obtained from the scikit-learn library \citep{scikit-learn}. The estimation step is performed using the default parameters of the \texttt{KernelDensity} class, which employs a Gaussian kernel with a bandwidth of 1 and the Euclidean metric to compute distances. The entropy for continuous attributes is then numerically approximated according to the following equation:

\begin{equation}
\begin{aligned}
H(I) = -\int_{-\infty}^{\infty} f(k) \log_2(f(k)) \, dk
\end{aligned}
\label{eq:shannon-continuous}
\end{equation}

However, the value can vary significantly depending on the chosen attribute. Continuous probability distributions may be more prone to issues such as extreme values or instability \citep{dubois2019generalization}. To address this concern, after calculating the entropy for a fixed number of segments along the x and y coordinates for a given attribute, we determine the quantile probability for each image contained within our dataset. This approach provides a measure that is flexible and comparable across different micro-CT images, allowing us to objectively assess heterogeneity relative to the entire dataset. By applying quantiles, we obtain a measure that is comparable across the chosen set of samples. This method can be used to identify which samples of a given type or with a shared characteristic exhibit greater heterogeneity or to compare samples from different types or regions, making it suitable for various reservoir characterization tasks.

\subsection*{Metrics and statistical assessment}

\subsubsection*{Comparison among the obtained rank results}

To evaluate the relationships between entropy ranks, which are derived by sorting the input samples based on their calculated entropy values given an attribute and a number of segments along the x and y dimensions, the Spearman correlation coefficient was employed. Spearman correlation is particularly suitable for rank-based data because it measures the strength and direction of the monotonic relationship between two variables without assuming a linear relationship, which makes it ideal when working with ranks. As a nonparametric measure, it does not require the input data to satisfy the assumption of normality, allowing for more flexibility when handling data that may not fit traditional parametric models. The Spearman correlation coefficient (\(\rho_s\)) is calculated as follows:

\begin{equation}
\rho_s = 1 - \frac{6 \sum d_i^2}{n(n^2 - 1)}
\end{equation}

where \(d_i\) is the difference between the ranks of the corresponding values of the two tested variables and \(n\) is the number of data points.

The corresponding values refer to paired observations derived from two variables \(A\) and \(B\) for the same entity. For example, \(R_A\) represents the entropy rank derived from attribute \(A\), and \(R_B\) represents the entropy rank derived from attribute \(B\), \(d_i = R_{A_i} - R_{B_i}\), where \(R_{A_i}\) and \(R_{B_i}\) represent the ranks of the \(i\)-th image. The formula is straightforward when no tied ranks are observed; however, in the presence of ties, a more general method involving the covariance of the ranks is applied \citep{dodge2008concise}.

\subsubsection*{Phi coefficient}

The Phi coefficient (\(\phi\)) is a measure of the association between two binary variables. The Phi coefficient is a special case of the Pearson correlation that is used when both examined variables are binary \citep{brusco2021comparison}. It quantifies the strength and direction of the association between two dichotomous outcomes, such as expert assessments of heterogeneity. This coefficient is calculated as follows:

\begin{equation}
\phi = \frac{n_{11}n_{00} - n_{10}n_{01}}{\sqrt{(n_{1\bullet}n_{0\bullet}n_{\bullet1}n_{\bullet0})}}
\end{equation}

where \(n_{11}, n_{00}, n_{10}, n_{01}\) are the frequencies of the contingency table cells, and \(n_{1\bullet}, n_{0\bullet}, n_{\bullet1}, n_{\bullet0}\) are the marginal sums, with \(n_{1\bullet}\) and \(n_{0\bullet}\) representing row sums and \(n_{\bullet1}\) and \(n_{\bullet0}\) representing column sums.

\subsubsection*{Rank-biserial correlation}

Rank-biserial correlation is a nonparametric association measure that evaluates the difference between the ranks two groups (e.g., expert classifications of heterogeneity or homogeneity) based on ordinal data, such as heterogeneity rankings \citep{cureton1956rank}. Unlike point-biserial correlation, which assumes normally distributed data, rank-biserial correlation makes no such assumption, making it more suitable when the conditions of normality and similar spreads (equal variances) are not satisfied. The rank-biserial correlation coefficient (\(\rho_{rb}\)) can be computed via the following formula:

\begin{equation}
\rho_{rb} = \frac{R_1}{n_1} - \frac{R_0}{n_0}
\end{equation}

where \(R_1\) and \(R_0\) are the sums of the ranks of the continuous variables for the two groups and \(n_1\) and \(n_0\) are the sizes of the two groups (e.g., the numbers of heterogeneous and homogeneous classifications, respectively).

\subsubsection*{Mann-Whitney U test}

The Mann-Whitney U (MWU) test is a nonparametric test used to assess whether a significant difference is present between the distributions of two groups \citep{mcknight2010mann}. It is commonly used in cases where the assumptions of a t test, i.e., normality within each group and homogeneity between the variances of the groups, are not met. The null hypothesis of the MWU test assumes that the two groups come from the same distribution, meaning that there is no significant difference between the groups. A low p value indicates that the null hypothesis can be rejected, suggesting that the distributions of the two groups are significantly different.

\subsubsection*{GLCM textural attributes}

The GLCM, introduced by \citet{haralick1973textural}, is a tool used to analyse texture by examining the spatial relationships in a 2-dimensional image. The GLCM analyses texture by counting how often pairs of pixels with specific grey levels appear together at a certain distance and direction from each other in an image. The matrix entries \(p(i,j)\) represent the normalized frequency at which a pixel with a grey level of \(i\) is adjacent to a pixel with a grey level of \(j\). Element \(p(i,j)\) in the GLCM represents the normalized frequency of such occurrences, yielding a pixel pair probability distribution for the image. This enables the extraction of attributes that convey meaningful aspects of the image, as shown in Table \ref{tab:glcm_attributes}.

\begin{table}[h!]
\centering
\caption{GLCM attributes with their formulas and descriptions}
\label{tab:glcm_attributes}
\begin{tabular}{@{} l l p{0.6\textwidth} @{}}
\toprule
\textbf{attribute} & \textbf{Formula} & \textbf{Description} \\
\midrule
\textbf{Energy} & $\sum_{i,j} p(i,j)^2$ & Measures the uniformity of texture. High energy values indicate homogeneous textures with few dominant pixel pairings. \\
\midrule
\textbf{Dissimilarity} & $\sum_{i,j} |i - j| p(i,j)$ & Quantifies the amount of variation or contrast between grey level pairs, with higher values indicating greater differences. \\
\midrule
\textbf{Homogeneity} & $\sum_{i,j} \frac{p(i,j)}{1 + |i - j|}$ & Measures the closeness of the distribution of elements in the GLCM to the diagonal, indicating how much similar pixel pairs dominate. \\
\bottomrule
\end{tabular}
\end{table}

\section{Results}\label{sec:results}

\begin{figure}[H]
    \centering
    \begin{tabular}{c}
        \includegraphics[width=0.8\textwidth]{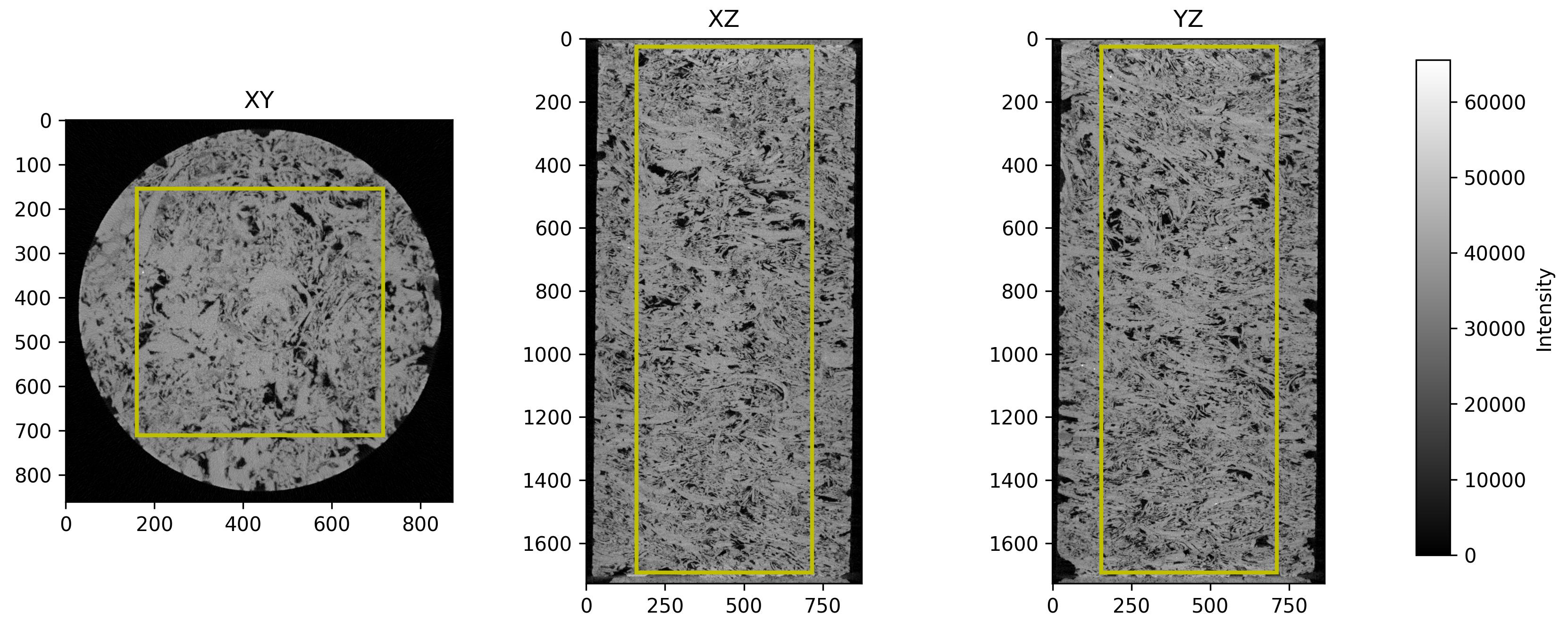} \\
        \small (a) A micro-CT image with a z-dimension that is larger than its x- and y-dimensions. \\
        \includegraphics[width=0.8\textwidth]{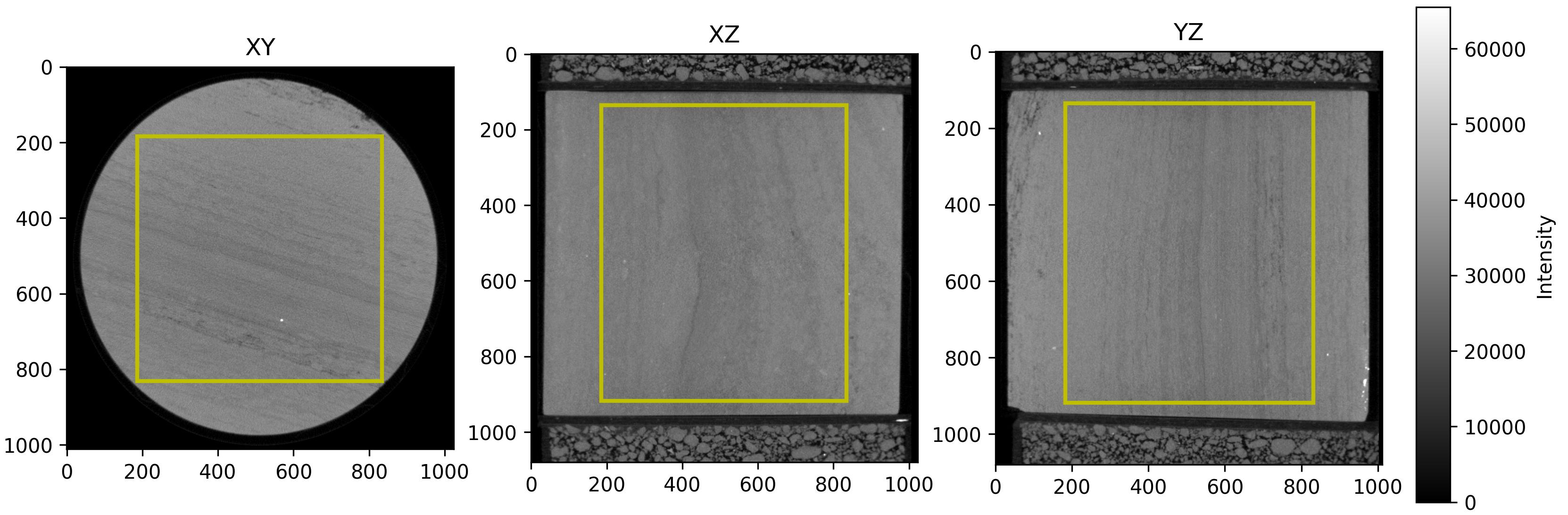} \\
        \small (b) A micro-CT image whose z-dimension is close in size to its x- and y-dimensions. \\
    \end{tabular}
    \caption{Central slices of micro-CT images derived from different samples. The VOI is delimited by yellow rectangles in each view.}
    \label{fig:example_cuts}
\end{figure}

The proposed method was tested on three discrete attributes (the minimum, maximum, and median) and five continuous attributes (the standard deviation, coefficient of variation, skewness, kurtosis, and mean). The x and y coordinates were divided into 2 to 10 segments, with the total number of subcubes varying depending on the dimensions of the given image. When 10 segments were used, the number of subcubes ranged from 1,000 to 4,200 across the 4,935 images. This large variation in the number of subcubes was due to differences between the z-dimension and the x- and y-dimensions, as exemplified by Figure \ref{fig:example_cuts}. The yellow rectangles delimit the VOI in the different views. The VOIs of the images, in terms of their dimensions, range from 275 to 5220 voxels. In Figure \ref{fig:example_cuts}a, the VOI has a larger z-dimension than its x- and y-dimensions, leading to the generation of more subcubes.

\subsection*{Behaviours under different attributes and numbers of divisions}

Not all attributes behaved as expected in relation to the intended purpose of the proposed method. The minimum, maximum, skewness, and kurtosis failed to capture meaningful relationships with heterogeneity based on the expert evaluation. The median and mean are measures of central tendency and are highly correlated, but for simplicity, we favoured the mean in the results. The standard deviation and coefficient of variation, which are inherent measures of variability, were tested, and their results aligned with the expert opinions, reinforcing the validity of the method when a favourable attribute was chosen. Kurtosis measures the "tailedness" of a distribution, with negative values indicating more tapered distributions and positive values suggesting heavier tails. Although kurtosis did not appear to correlate with heterogeneity in our initial visual analysis, it was included to emphasize the value of choosing the correct attributes by contrasting it with those that produced better results. Therefore, we present the standard deviation, coefficient of variation, mean, and kurtosis in the results.

\begin{figure}[H]
    \centering
    \begin{tabular}{cc}
        \includegraphics[width=0.45\textwidth]{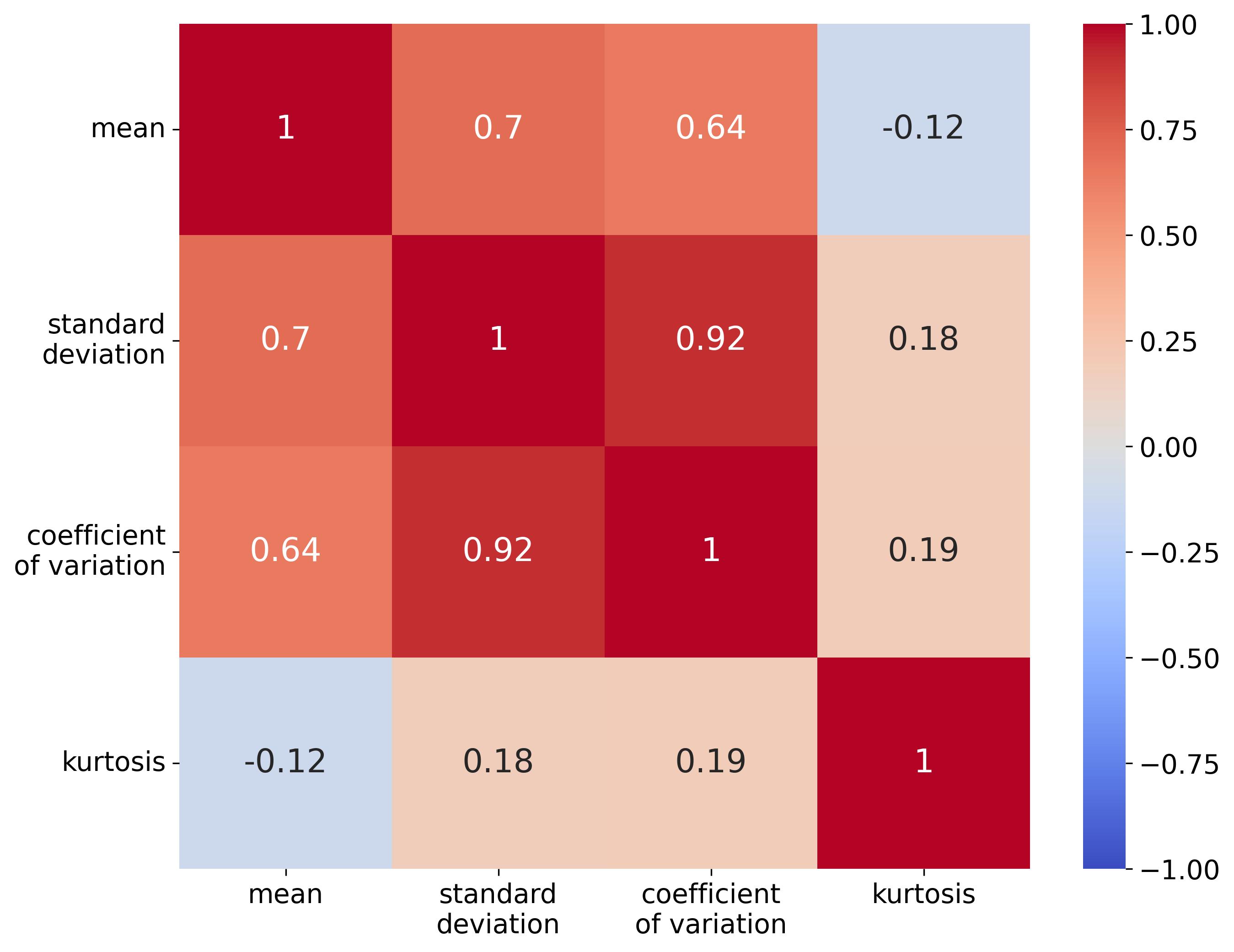} &
        \includegraphics[width=0.45\textwidth]{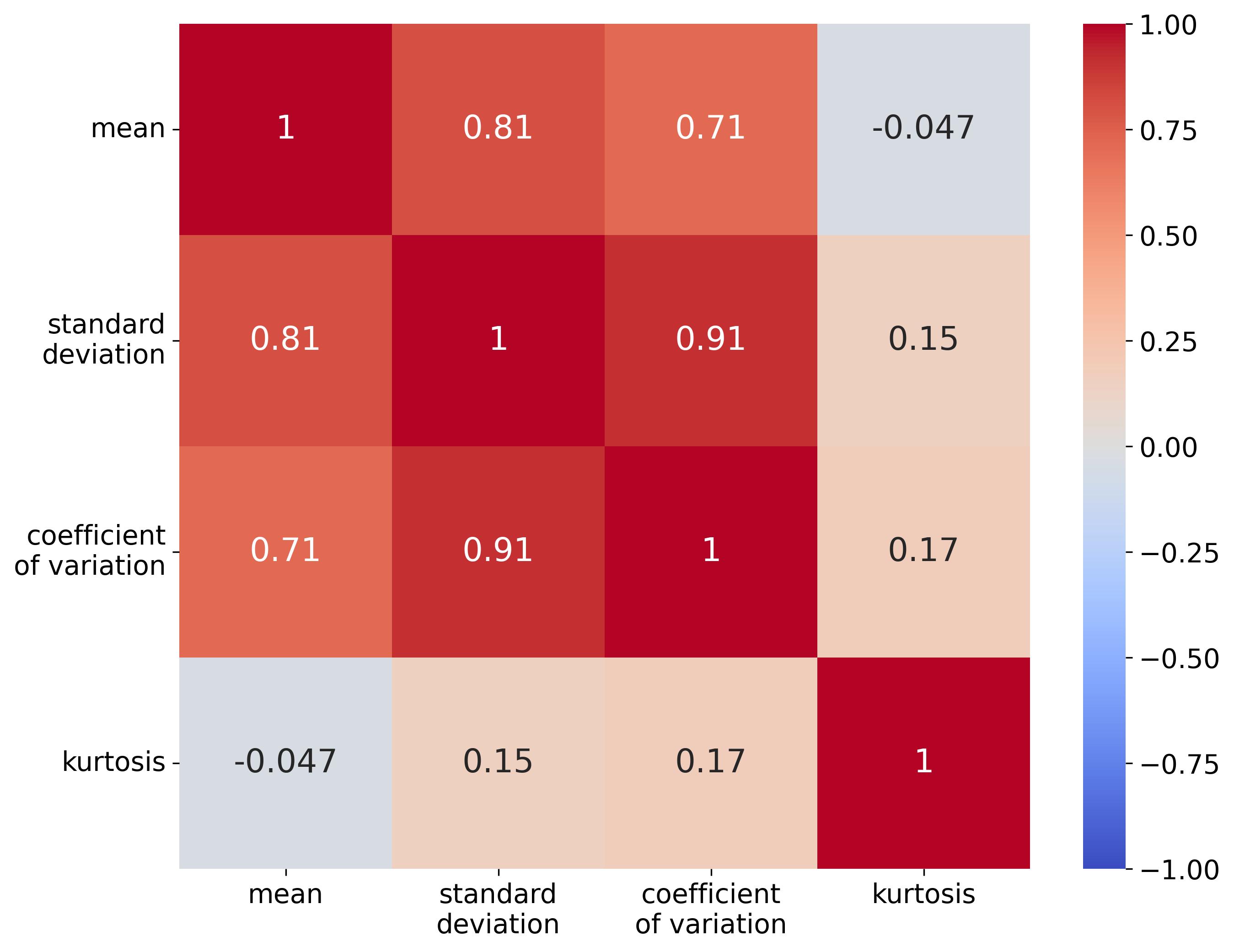} \\
        \parbox{0.45\textwidth}{\centering \small (a) Spearman correlation heatmaps between different attributes: 5 segments.} &
        \parbox{0.45\textwidth}{\centering \small (b) Spearman correlation heatmaps between different attributes: 10 segments.} \\
        \includegraphics[width=0.45\textwidth]{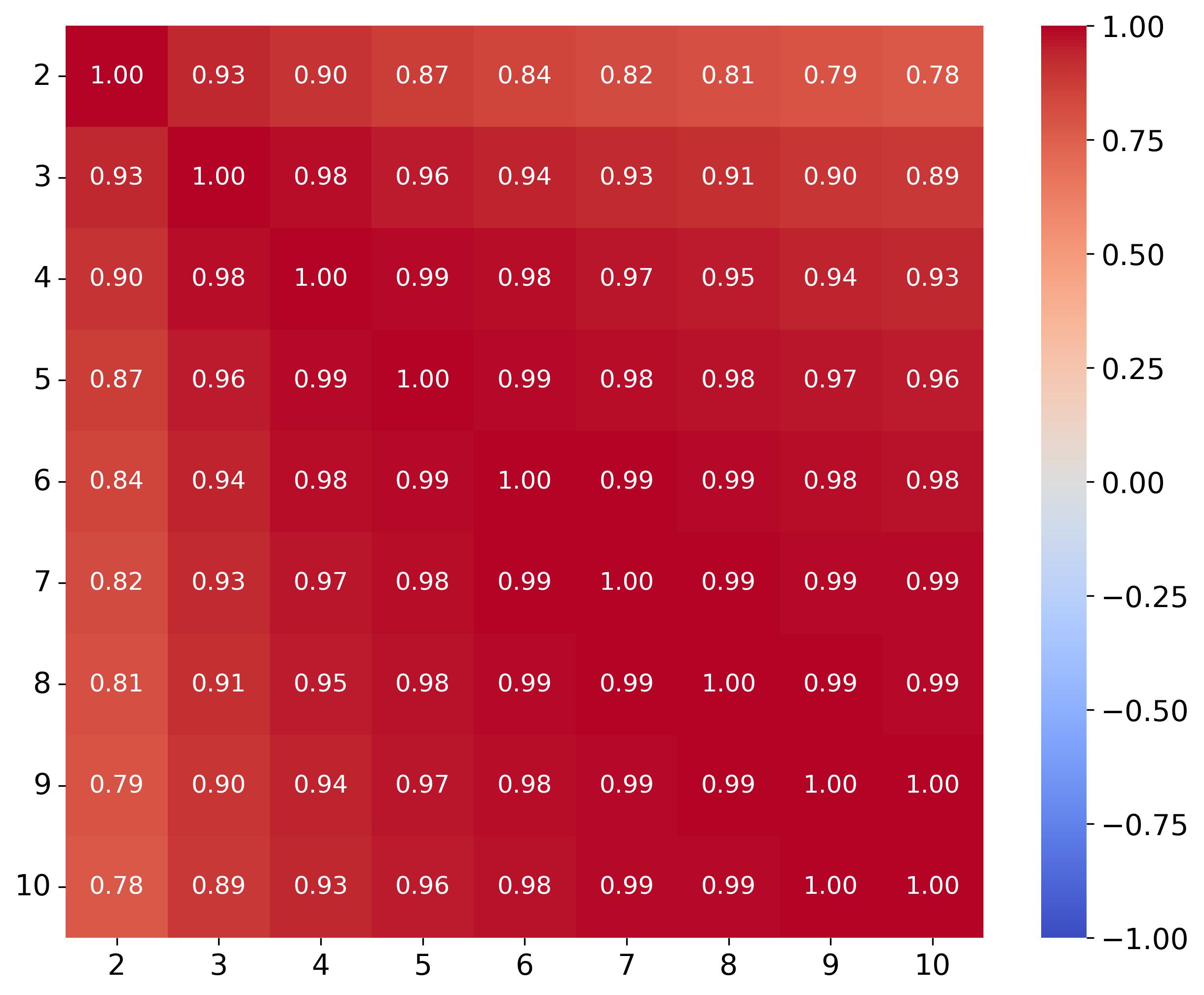} &
        \includegraphics[width=0.45\textwidth]{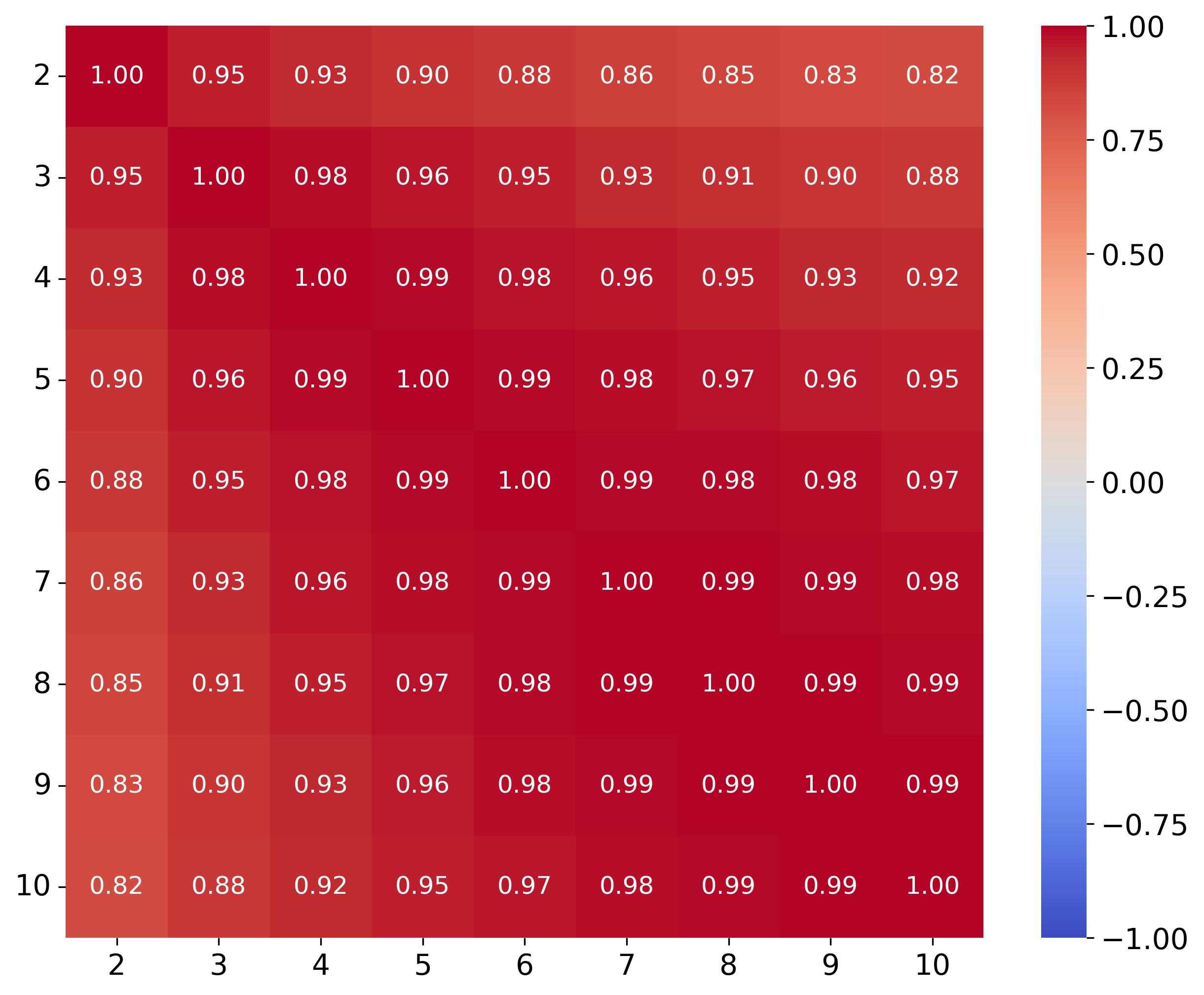} \\
        \parbox{0.45\textwidth}{\centering \small (c) Spearman correlations for the mean attribute across different division schemes.} &
        \parbox{0.45\textwidth}{\centering \small (d) Spearman correlations for the standard deviation attribute across different division schemes.} \\
    \end{tabular}
    \caption{Spearman correlation matrices produced for the entropy ranks of different attributes and division schemes with respect to the x and y coordinates. (a) 5 segments, (b) 10 segments, (c) Mean attribute, (d) Standard deviation attribute.}
    \label{fig:spearman}
\end{figure}

Spearman correlation was applied to explore the relationships between the measures observed across different attributes and the number of divisions, as shown in Figure \ref{fig:spearman}. Figures \ref{fig:spearman}a and \ref{fig:spearman}b present the correlation matrices between different attributes and the entropy ranks obtained via 5 and 10 segments along the x and y coordinates. The results revealed that the standard deviation and coefficient of variation strongly correlate with each other. The mean displayed slightly weaker correlations with them. In contrast, kurtosis exhibited consistently low correlations with the other selected attributes, indicating that the ranks it produced diverged significantly from those of the other attributes. The overall trends were consistent across both division configurations.

Figures \ref{fig:spearman}c and \ref{fig:spearman}d illustrate the Spearman correlation heat maps produced for the mean and standard deviation attributes in different numbers of segments. Both heat maps demonstrate strong positive correlations, starting from a minimum value of 0.78. Notably, the correlations involving the 2-segment case were slightly weaker, especially compared with those produced under higher segment counts, suggesting that the ranks derived from only 2 segments were less consistent with those obtained from finer subdivision schemes. Overall, the heat maps indicate that while the levels of granularity introduced by different division schemes had an impact, the ranks remained strongly correlated throughout the procedure. Therefore, for the sake of simplicity, all subsequent results focus on the 5-segment case for the x and y coordinates case. Additionally, in some instances, specific attributes are selected for further analysis.

\subsection*{Visualization of the top and bottom 5 images}

\begin{figure}[H]
    \centering
    \includegraphics[width=\textwidth]{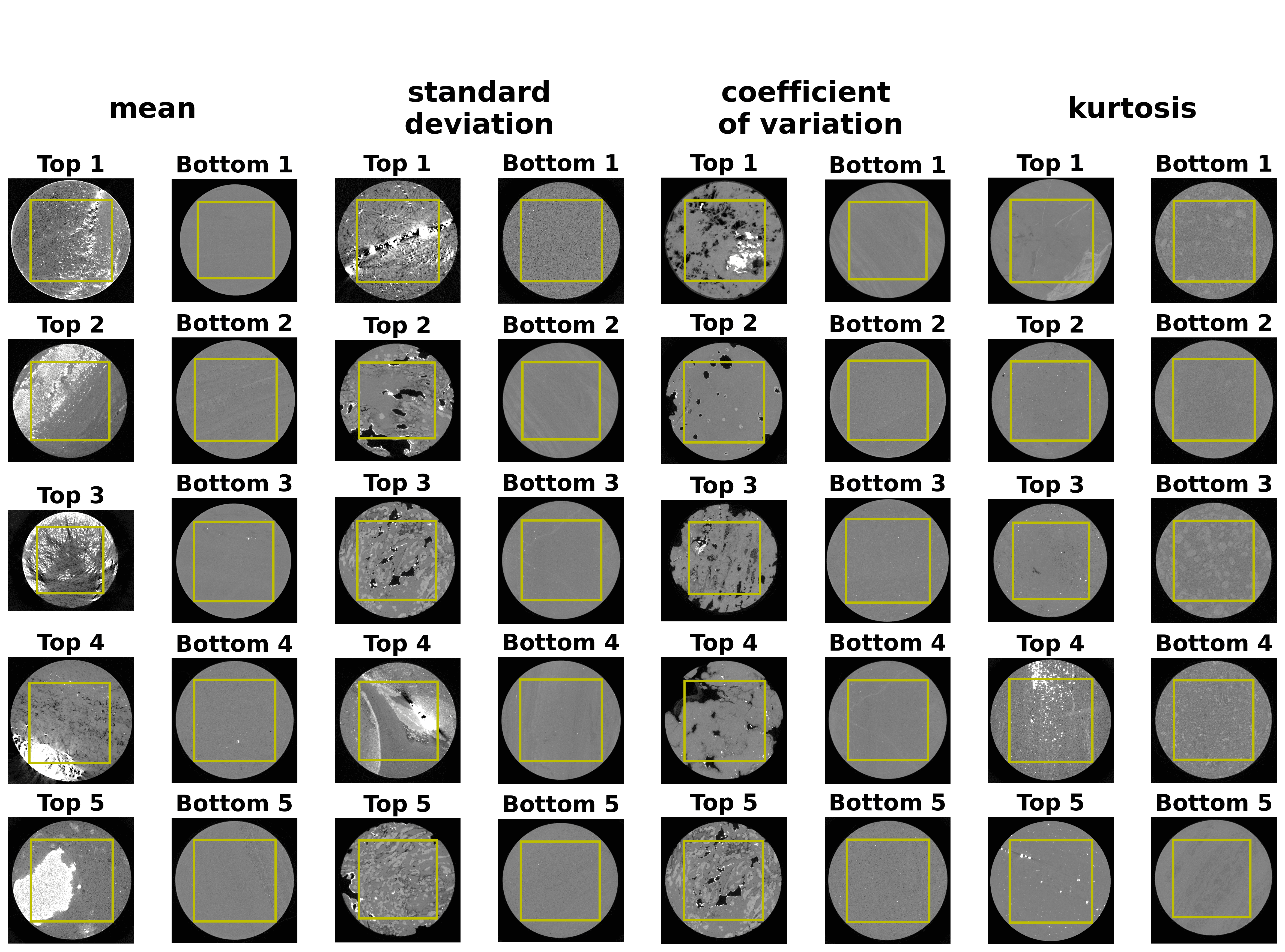}
    \caption{Horizontal cross-section centre slices of the images with the highest and lowest entropy values for the selected attributes. The yellow boxes delimit the corresponding area within each VOI.}
    \label{fig:slices_top_bottom}
\end{figure}

To better understand how the above attributes impact the heterogeneity coefficient, Figure \ref{fig:slices_top_bottom} displays the central cross-sectional slices of the images with the five highest and lowest entropy values for each attribute. Ideally, these slices should represent the most heterogeneous and most homogeneous samples, respectively. However, this pattern is not consistently observed for the kurtosis attribute, which displays a completely different behaviour, as expected from the results presented in Figure \ref{fig:spearman}. In contrast, the results obtained for the CV, mean and standard deviation attributes align well with the proposed hypothesis. When these attributes are selected, a clear distinction emerges between the high- and low-entropy micro-CT scans, with the high-entropy group clearly appearing more heterogeneous.

\begin{figure}[H]
    \centering
    \includegraphics[width=\textwidth]{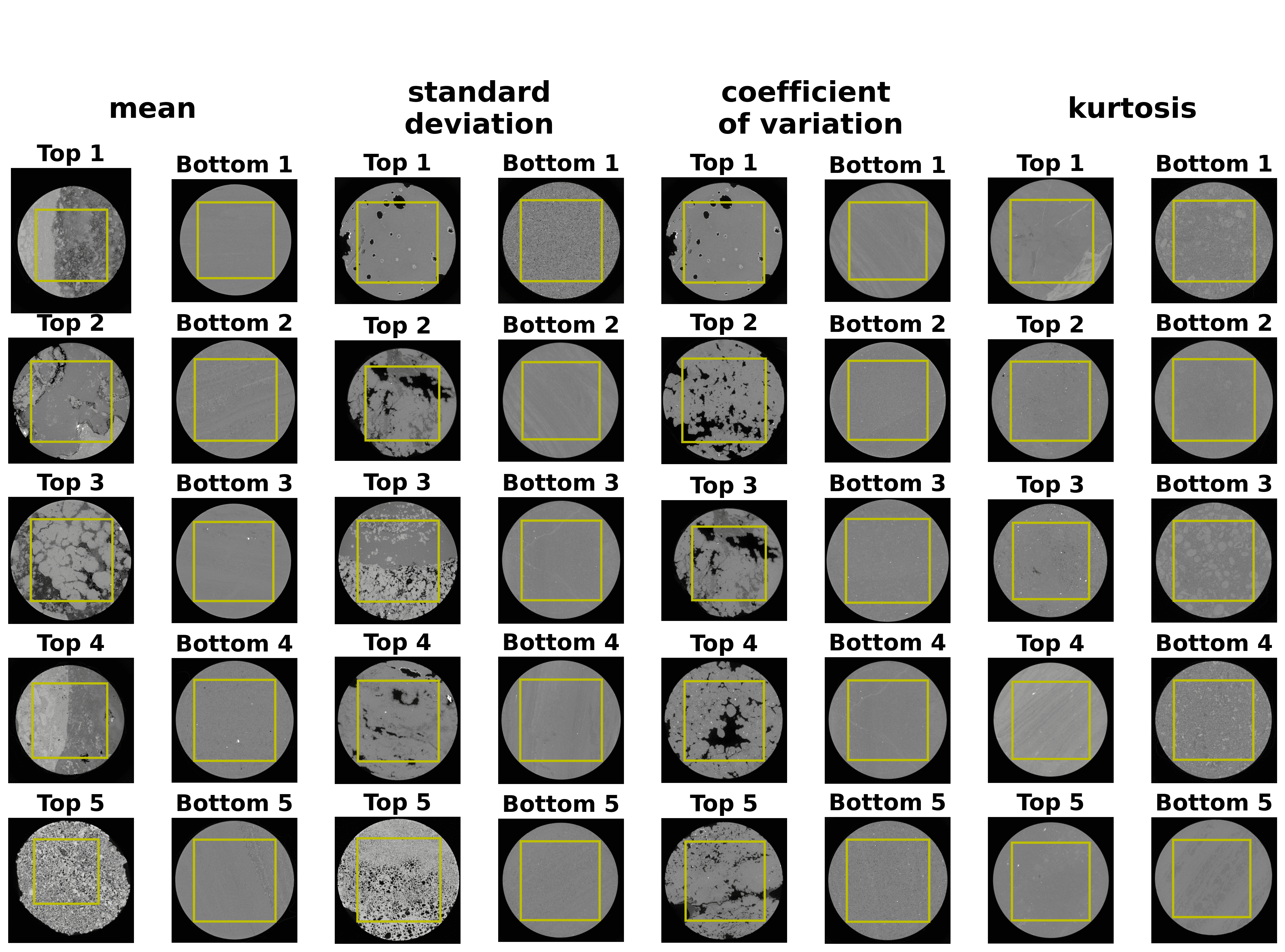}
    \caption{Horizontal cross-section centre slices of the images with the highest and lowest entropy values after a high-density filter was applied for the selected attributes. The yellow boxes delimit the corresponding area within each VOI.}
    \label{fig:slices_top_bottom_no_high_density}
\end{figure}

The dataset presents some attributes that might cause incorrect predictions, such as different luminous conditions that are produced due to the polychromatic nature of X-ray beams and image artefacts, which can be caused by high-density materials. The grey levels in a reconstructed CT image represent X-ray attenuation through the material, causing higher-density phases to appear brighter \citep{singhal2013micro}. To minimize the interference caused by artefacts and high-density regions, a filter was applied to remove all images with more than $0.1\%$ of the voxels possessing intensity values higher than 60,000. Figure \ref{fig:slices_top_bottom_no_high_density} displays the images with the highest and lowest entropy values for each attribute in this scenario. The results remain consistent, suggesting that even though high-density materials can produce outliers leading to elevated entropy values and ranks, excluding these images, the distinction between the high- and low-entropy micro-CT scans remains evident.

\subsection*{Analysis of selected Aptian-age formations}

We selected two Aptian-age formations \citep{milani2007bacias}, each representing the final sedimentary layer deposited before the onset of salt accumulation in two distinct coastal sedimentary basins. The data were grouped by facies to enable a more controlled assessment of heterogeneity by focusing on samples within specific facies groups. This approach allows experts to evaluate heterogeneity within well-defined geological contexts.

\begin{figure}[H]
    \centering
    \includegraphics[width=\textwidth]{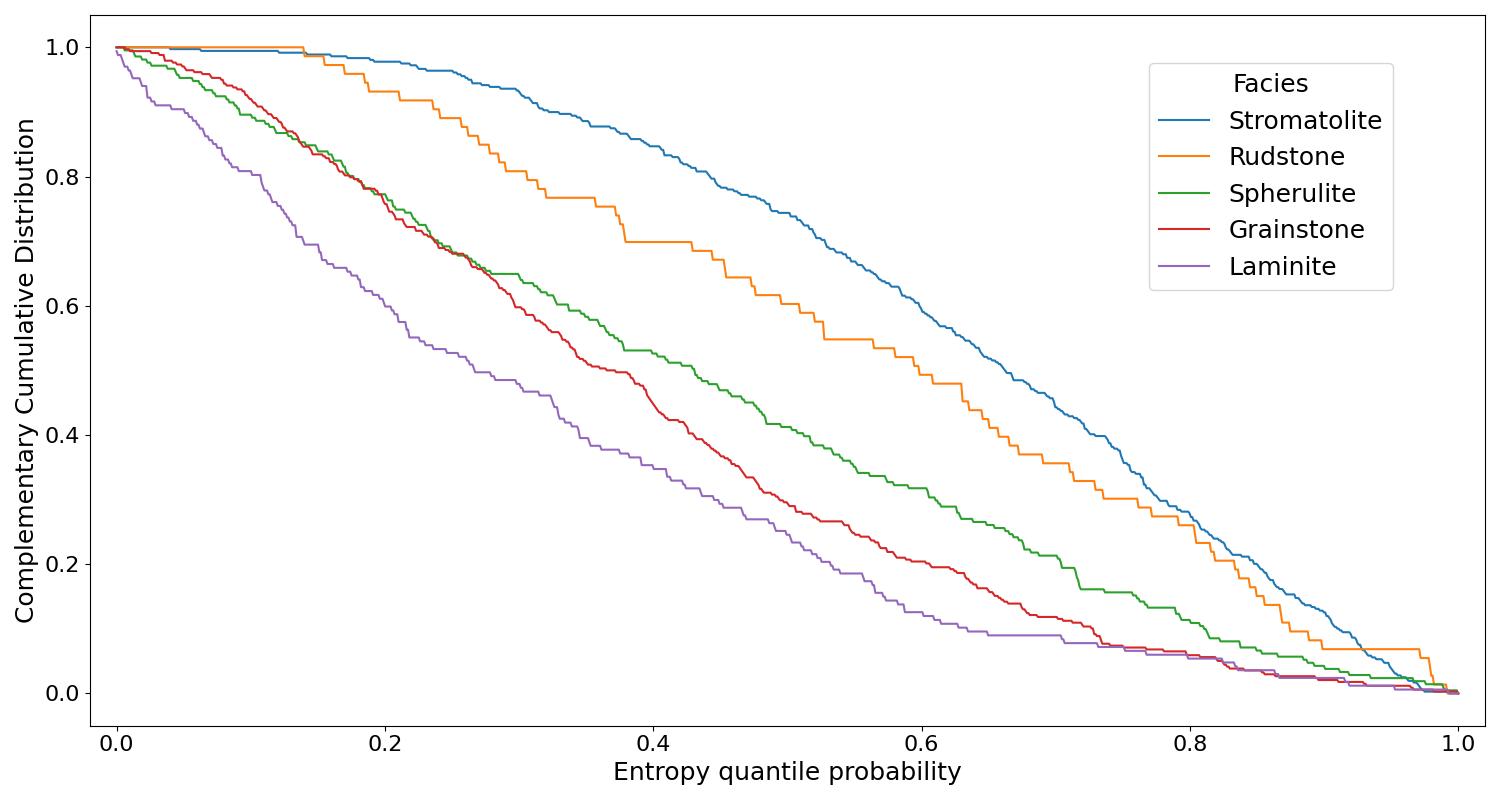}
    \caption{Cumulative distribution plot of the textural heterogeneity measure across five facies from the final Aptian-age formation of the two coastal sedimentary basins. The x-axis represents our measure, while the y-axis shows the cumulative distribution percentage for each facies.}

    \label{fig:aptian_age_cumplot}
\end{figure}

The plot in Figure \ref{fig:aptian_age_cumplot}  illustrates the complementary cumulative distribution of our measurements across five facies from these formations. In general, rudstone and stromatolite exhibit the highest levels of heterogeneity, as indicated by their slower decline in the lower range of the complementary cumulative distribution curves. This pattern reflects a concentration of values at higher levels of the textural heterogeneity measure. In contrast, laminite displays a more textural homogeneous profile, with a complementary cumulative distribution curve that declines more steeply, reflecting a lower proportion of values at higher heterogeneity levels. 

In the case of stromatolites and rudstones, their pore systems are characterized by a range of pore sizes, some larger than the rock framework itself, often resulting from dissolution processes. Laminite, while also exhibiting a complex pore structure when observed at higher resolutions, generally displays more uniform pore sizes, contributing to a more homogeneous profile in the textural heterogeneity measure. However, it is important to note that this result is not the expected following the geological concept of heterogeneity, since grainstones were expected to be more homogeneous than laminites. Experts identified that our measure of textural heterogeneity reflects variations arising from lithological changes, diagnostic minerals, porous structures, and potential artifacts, such as barite residue from incomplete sample cleaning. While the measure does not always align with the geological concept of heterogeneity, it remains a valuable tool for efficient analyses, offering insights into textural variations.

\begin{figure}[H]
    \centering
    \begin{tabular}{cc}
        \includegraphics[width=0.45\textwidth]{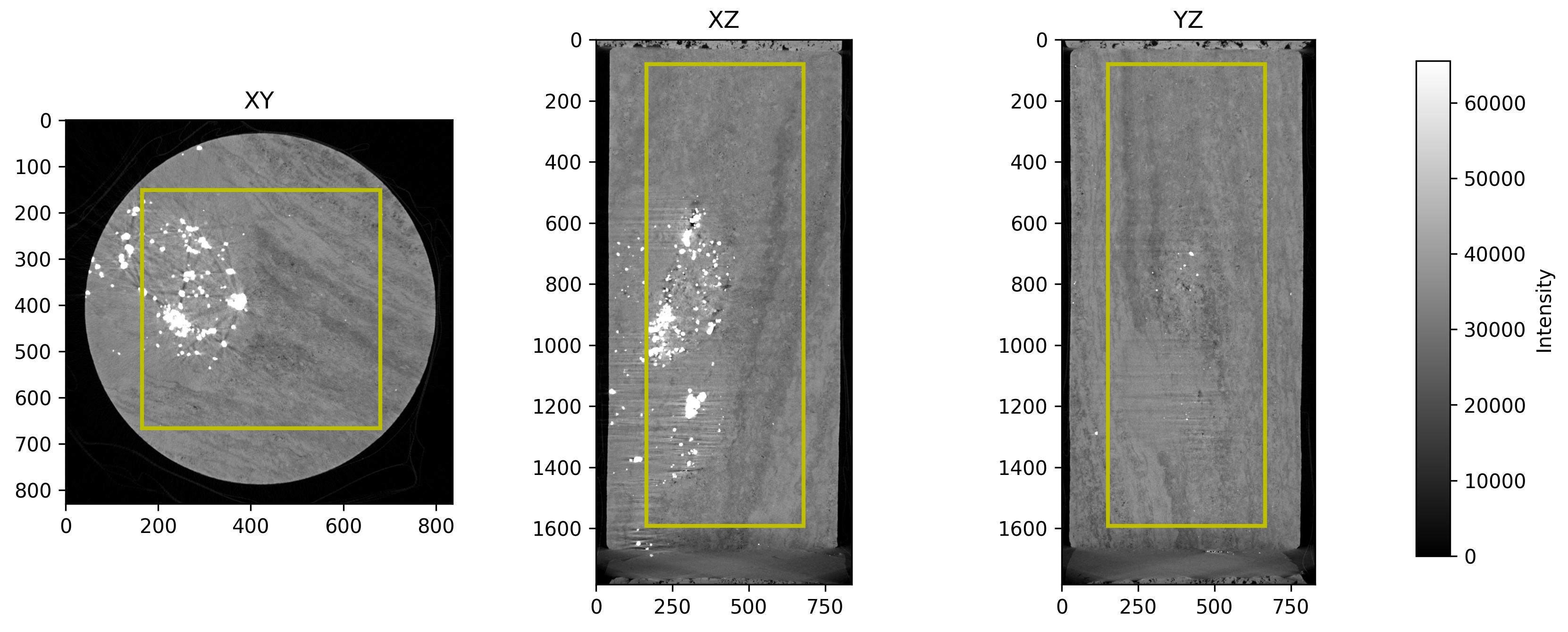} & 
        \includegraphics[width=0.45\textwidth]{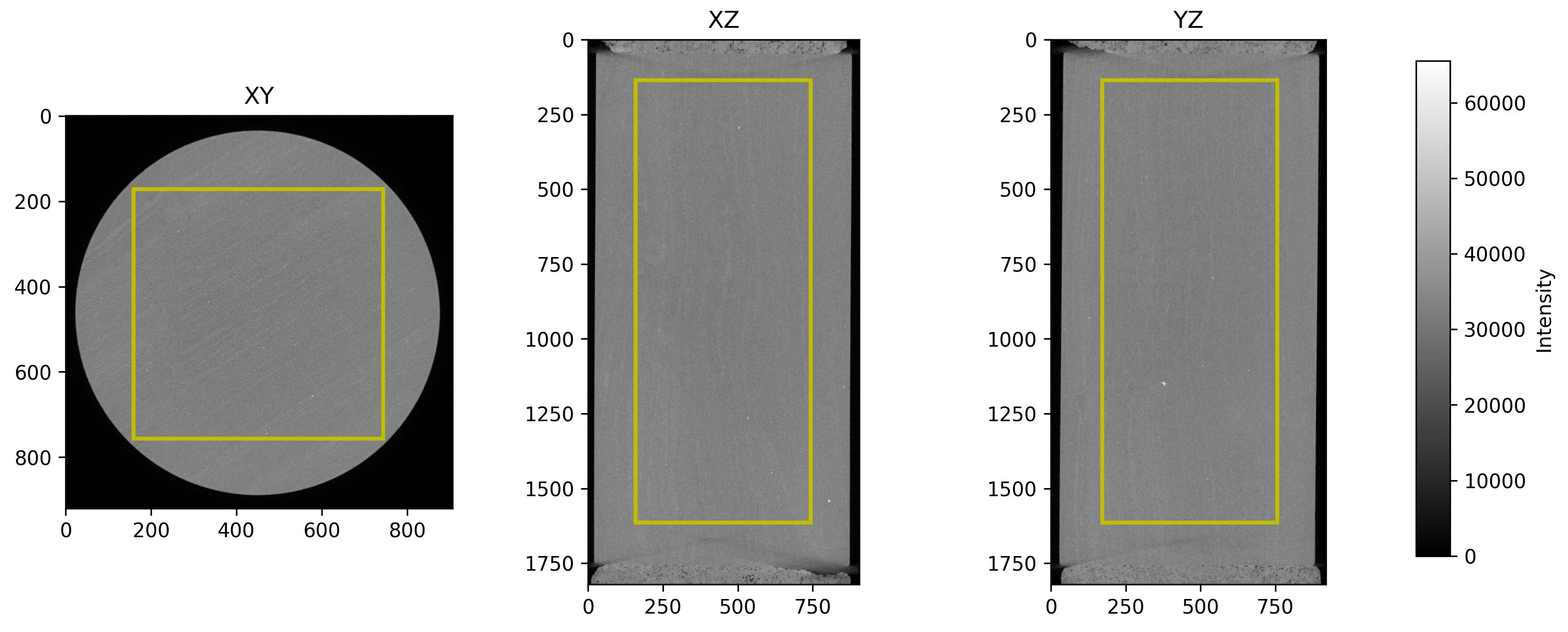} \\
        \small (a) Laminite - high heterogeneity & 
        \small (b) Laminite - low heterogeneity \\
        \includegraphics[width=0.45\textwidth]{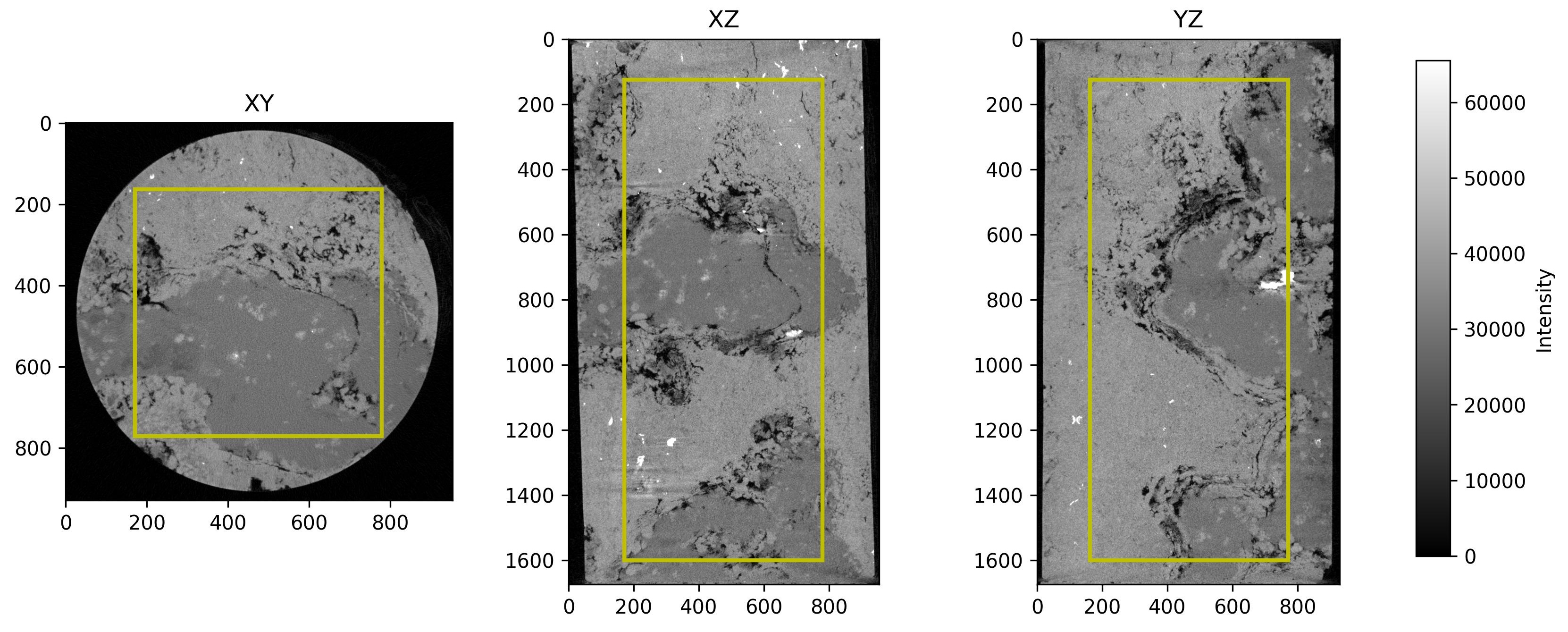} & 
        \includegraphics[width=0.45\textwidth]{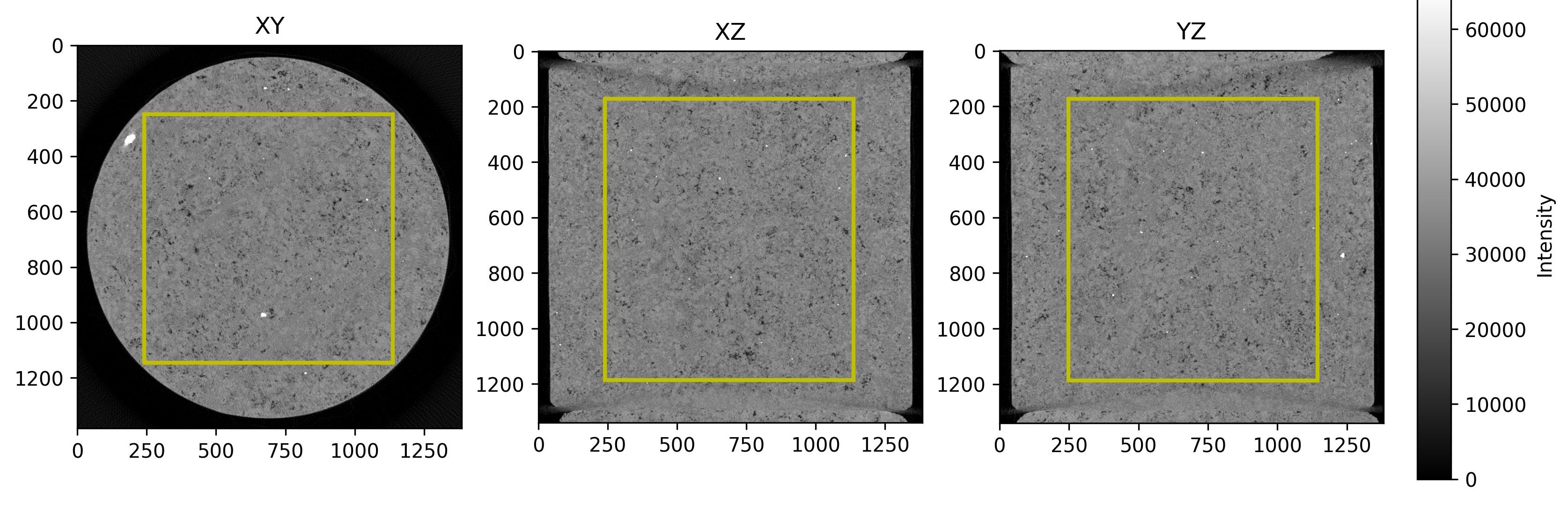} \\
        \small (c) Stromatolite - high heterogeneity & 
        \small (d) Stromatolite - low heterogeneity \\
        \includegraphics[width=0.45\textwidth]{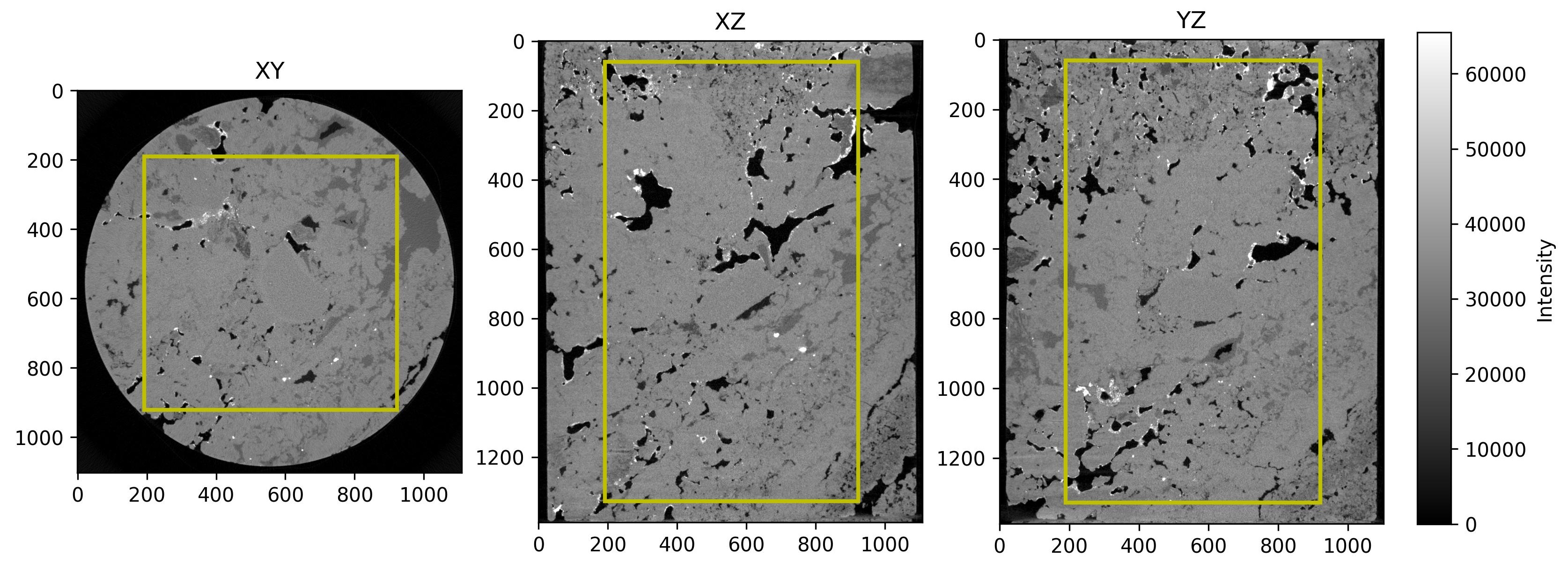} & 
        \includegraphics[width=0.45\textwidth]{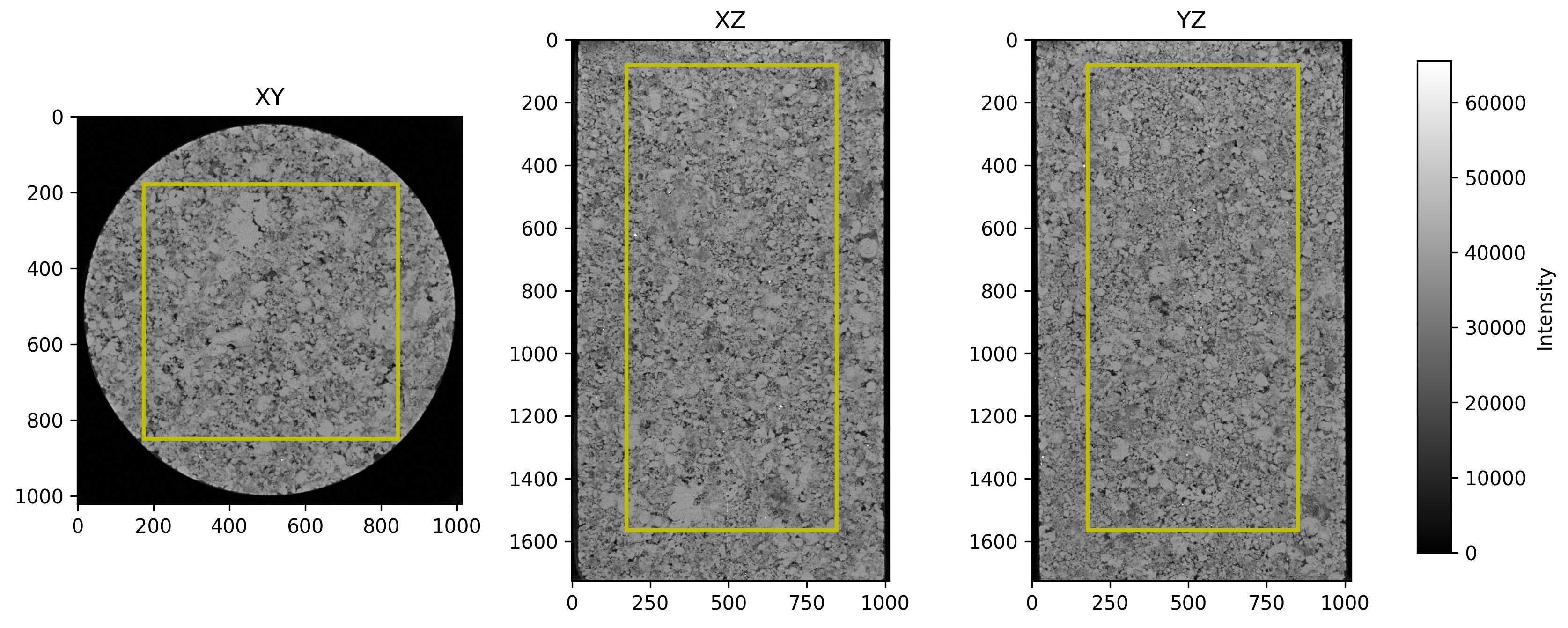} \\
        \small (e) Rudstone - high heterogeneity & 
        \small (f) Rudstone - low heterogeneity \\
    \end{tabular}
    \caption{Examples of micro-CT images with high and low textural heterogeneity measurements for each facies. (a) Laminite with high heterogeneity, (b) Laminite with low heterogeneity, (c) Stromatolite with high heterogeneity, (d) Stromatolite with low heterogeneity, (e) Rudstone with high heterogeneity, (f) Rudstone with low heterogeneity.}
    \label{fig:aptian_examples}
\end{figure}

Figure \ref{fig:aptian_examples} presents micro-CT images displaying both high and low textural heterogeneity measures for facies with varying levels of heterogeneity: laminite, which generally exhibits lower heterogeneity, and rudstone and stromatolite, which show higher heterogeneity. In Figure \ref{fig:aptian_examples}a, the high heterogeneity ranking of the laminite sample may be attributed to the presence of high-density minerals. This is an exception, as most laminite samples generally fall on the lower end of the measure. Figure \ref{fig:aptian_examples}c illustrates stromatolite with notable lithological changes, characterized by distinct structures and compositional variability within the sample. Meanwhile, Figure \ref{fig:aptian_examples}e depicts a rudstone sample in which heterogeneity arises primarily from a complex and irregular pore structure.

\subsection*{Comparison with an expert heterogeneity assessments}

\begin{figure}[H]
    \centering
    \begin{tabular}{c}
        \includegraphics[width=0.8\textwidth]{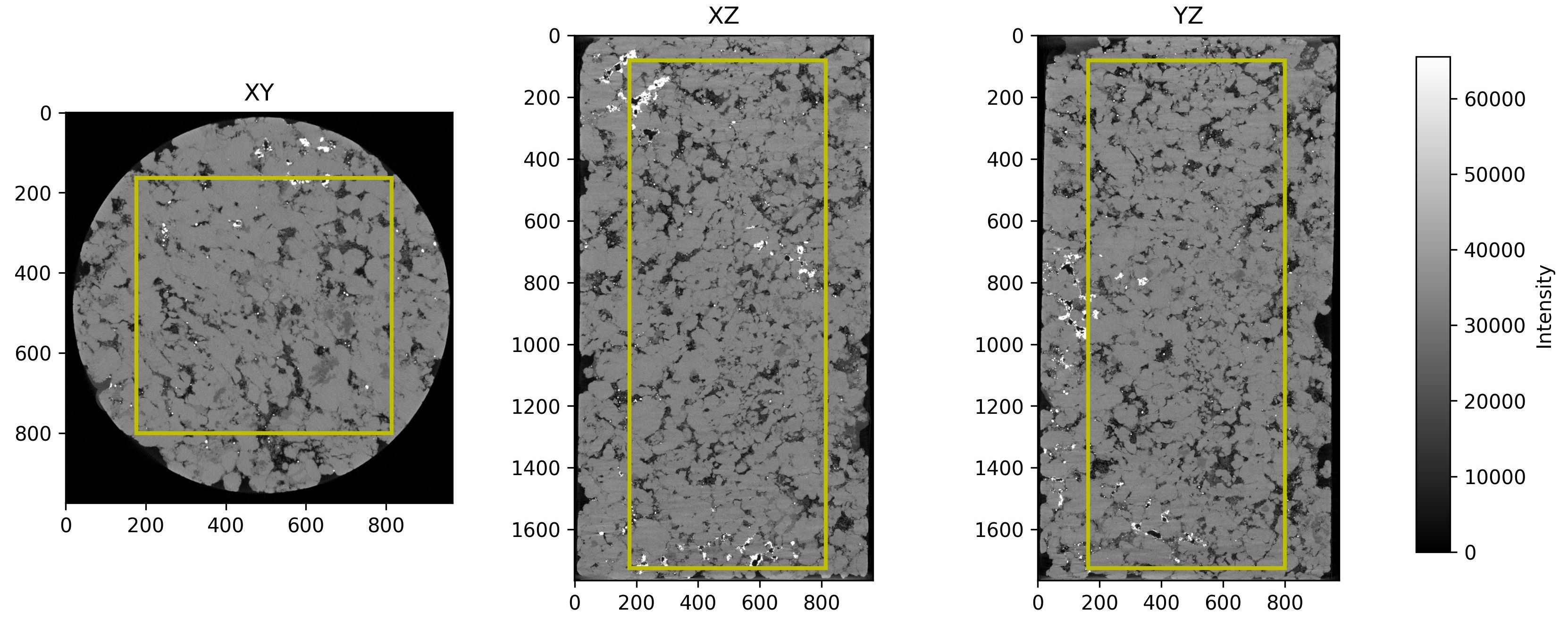} \\
        \parbox{0.8\textwidth}{\centering \small (a) Central slices of a micro-CT image from a sample unanimously labeled as homogeneous by all four experts.} \\
        \includegraphics[width=0.8\textwidth]{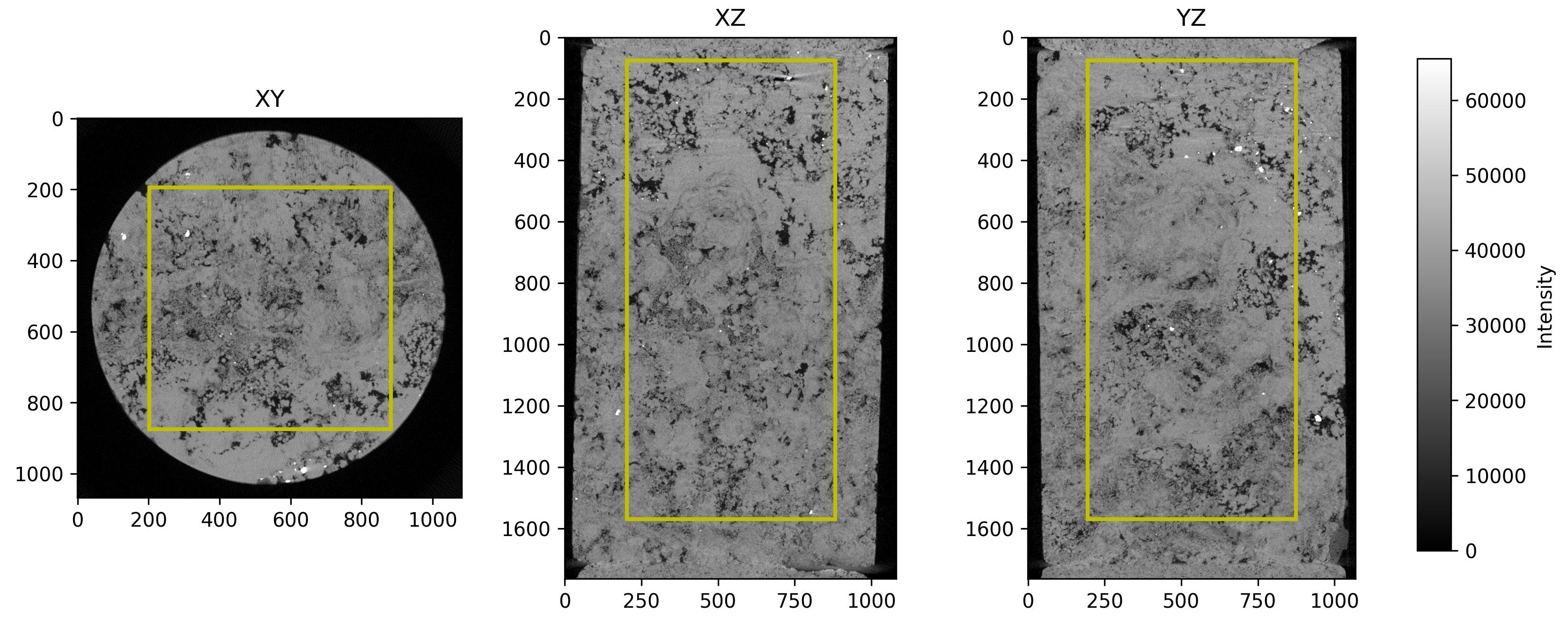} \\
        \parbox{0.8\textwidth}{\centering \small (b) Central slices of a micro-CT image from a sample unanimously labeled as heterogeneous by all four experts.} \\
        \includegraphics[width=0.8\textwidth]{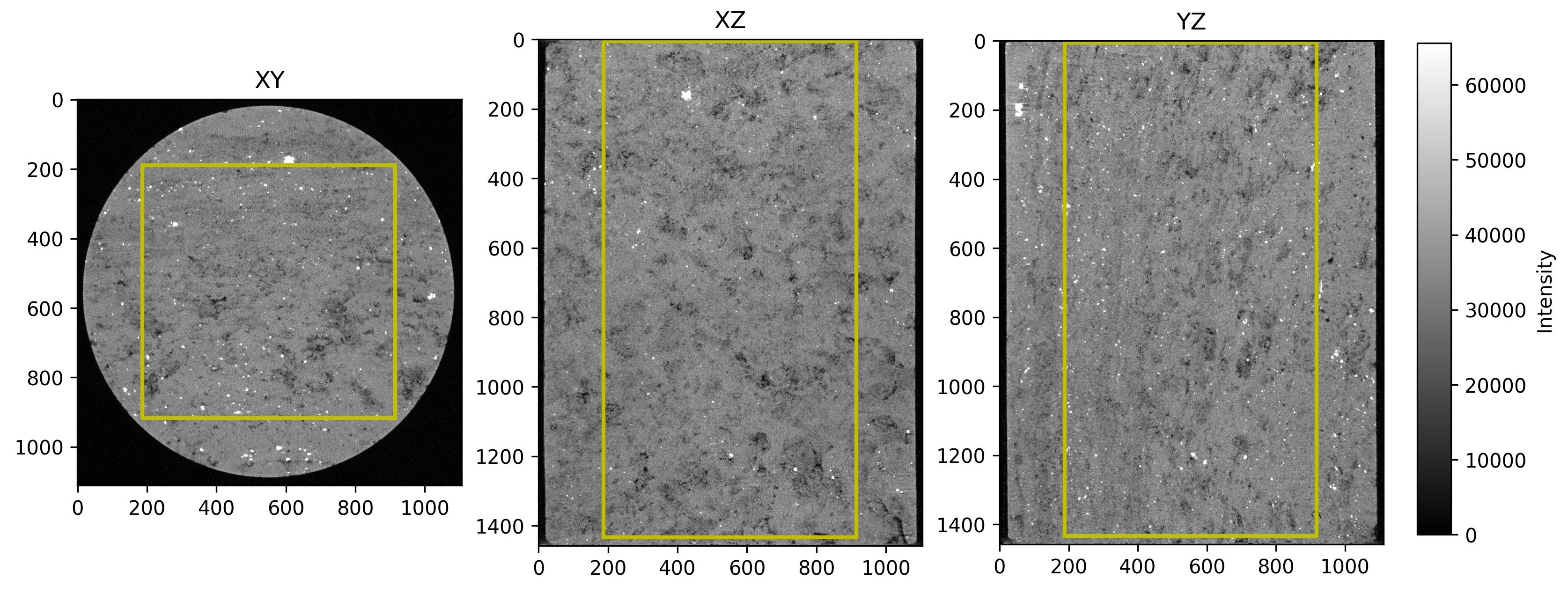} \\
        \parbox{0.8\textwidth}{\centering \small (c) Central slices of a micro-CT image from a sample with conflicting labels: one expert assessed it as homogeneous, and another as heterogeneous.} \\
    \end{tabular}
    \caption{Central slices of micro-CT images showcasing examples of expert-labeled samples: homogeneous, heterogeneous, and conflicting assessments. (a) Homogeneous, (b) Heterogeneous, (c) Conflicting.}
    \label{fig:examples_annotations}
\end{figure}

Out of the 4,744 samples contained in the dataset, 175 were labelled by at least one specialist as either heterogeneous or not. Since most samples are carbonates derived from a presalt region, they inherently exhibited some degree of heterogeneity, which was considered by the experts during the assessment process. The number of samples each expert evaluated varied, as shown in Table \ref{tab:label_count}. Examples of micro-CT images showcasing expert-labeled samples are shown in Figure \ref{fig:examples_annotations}. Figure \ref{fig:examples_annotations}a presents a sample unanimously identified as homogeneous, Figure \ref{fig:examples_annotations}b displays a sample unanimously labeled as heterogeneous, and Figure \ref{fig:examples_annotations}c represents a case with conflicting labels, where one expert assessed the sample as homogeneous and another as heterogeneous.

\begin{table}[h!]
\centering
\begin{minipage}{0.6\textwidth}
\centering
\begin{tabular}{l c c c}
\toprule
 & \textbf{Homogeneous} & \textbf{Heterogeneous} & \textbf{Total} \\
\midrule
\textbf{Expert 1} & 38 & 122 & 160 \\
\textbf{Expert 2} & 50 & 48  & 98  \\
\textbf{Expert 3} & 9  & 19  & 28  \\
\textbf{Expert 4} & 16 & 26  & 42  \\
\bottomrule
\end{tabular}
\end{minipage}%
\hspace{0.4cm}
\begin{minipage}{0.35\textwidth}
\caption{Numbers of homogeneous and heterogeneous assessments performed by each specialist.}
\label{tab:label_count}
\end{minipage}
\end{table}

\begin{figure}[H]
    \centering
    \includegraphics[width=\textwidth]{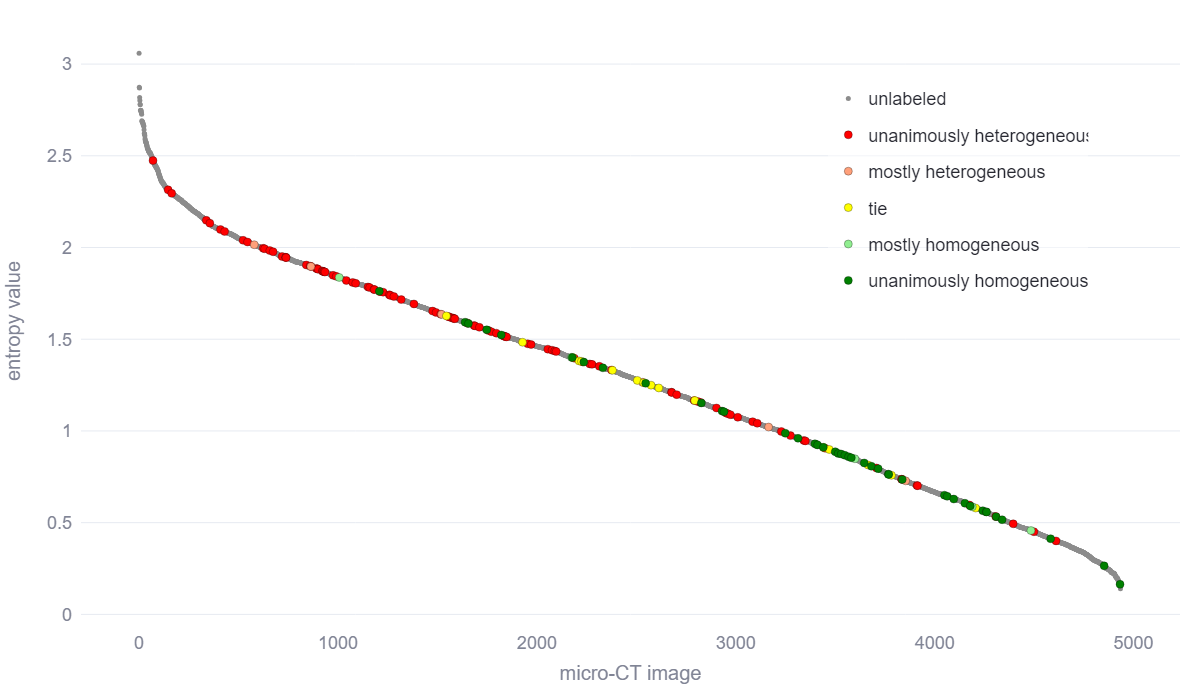}
    \caption{Micro-CT images ordered by their entropy values calculated from the standard deviation attribute. The grey dots represent images for which the corresponding samples were not labelled by experts. The coloured dots represent the expert assessments: dark green signifies unanimous agreement regarding a heterogeneous sample, light green indicates a majority agreement, yellow represents a tie between the numbers of heterogeneous and homogeneous labels, light red indicates a majority agreement regarding a homogeneous sample, and dark red signifies unanimous agreement concerning a homogeneous sample.}
    \label{fig:annot_entropy}
\end{figure}

Figure \ref{fig:annot_entropy} presents the entropy values produced for the standard deviation attribute in descending order. The samples are marked with dots possessing different colours based on the basis of the annotations provided by the specialists: green for heterogeneous samples and red for homogeneous samples. The plot clearly shows that the samples identified as heterogeneous tend to cluster on the left side, where the entropy values are higher, whereas homogeneous samples are more prevalent on the right, where the entropy values are lower. The samples for which the experts diverged are generally located in the middle of the entropy value range. This distribution demonstrates a positive correlation between higher entropy values and the heterogeneity assessments provided by specialists.

\begin{figure}[H]
    \centering
    \includegraphics[width=\textwidth]{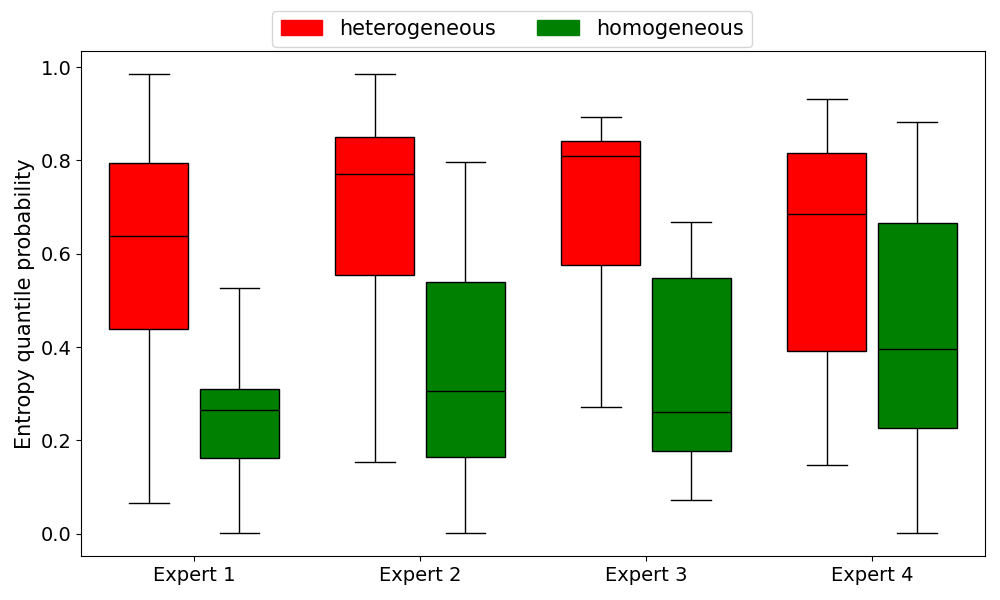}
    \caption{Box plots of the entropy quantile probabilities produced by each expert and the heterogeneity assessment results; attribute: standard deviation.}
    \label{fig:box_experts}
\end{figure}

Figure \ref{fig:box_experts} shows the distribution of the entropy quantile probabilities derived from the standard deviation attribute for the samples assessed by the experts, where red boxes represent images of samples labelled as heterogeneous samples and green boxes represent those labelled as being homogeneous. Across all the experts, heterogeneous samples led to images with higher entropy values, whereas homogeneous samples exhibited the opposite trend. This finding indicates that our method correlates positively with structurally heterogeneous samples according to the expert assessment results.

Table \ref{tab:phi_contingency} presents a comparison among the binary assessments of heterogeneity or homogeneity provided by different pairs of experts. The data indicate the levels of agreement among the experts regarding the characterizations of rock samples. The $\phi$ coefficient points to strong positive relationships in all the cases. This correlation was expected given that the expert opinions tended to converge when the images exhibited clear indicators of heterogeneity or homogeneity. However, the degree of disagreement between the experts was considerable. For example, Experts 1 and 2 agreed on 63 samples but disagreed on 20 samples, include 18 cases where Expert 1 deemed a sample to be heterogeneous while Expert 2 did not and 2 cases where Expert 2 considered a sample to be heterogeneous while Expert 1 did not. This level of disagreement underscores the subjectivity of the task and why automation is an ongoing challenge.

\begin{table}[h!]
\centering
\caption{Phi coefficient and contingency table values produced during the comparison among the heterogeneity assessments performed by the specialists. "True-True" represents the number of samples that both experts labelled as heterogeneous, "True-False" represents samples labelled as heterogeneous by the first expert and as homogeneous by the second expert, "False-True" represents samples labelled as homogeneous by the first expert and as heterogeneous by the second expert, and "False-False" represents samples that both experts labelled as homogeneous.}
\label{tab:phi_contingency}
\begin{tabular}{p{1.5cm} p{1.5cm} p{1cm} p{1.5cm} p{1.5cm} p{1.5cm} p{1.5cm}}
\toprule
 &  & \boldmath{$\phi$} & \textbf{True-True} & \textbf{True-False} & \textbf{False-True} & \textbf{False-False} \\
\midrule
\textbf{Expert 1} & \textbf{Expert 2} & 0.54 & 43 & 18 & 2 & 20 \\
         & \textbf{Expert 3} & 0.74 & 15 & 1  & 1 & 4  \\
         & \textbf{Expert 4} & 0.64 & 23 & 3  & 2 & 7  \\
\midrule
\textbf{Expert 2} & \textbf{Expert 3} & 0.64 & 13 & 0  & 6 & 9  \\
         & \textbf{Expert 4} & 0.53 & 21 & 4  & 5 & 11 \\
\midrule
\textbf{Expert 3} & \textbf{Expert 4} & 0.66 & 11 & 4  & 0 & 6  \\
\bottomrule
\end{tabular}
\end{table}


The association between the samples labelled as heterogeneous by experts and the higher entropy values (and consequently higher ranks) anticipated in Figure \ref{fig:box_experts} is confirmed by the data presented in Table \ref{tab:rankbiserial}. This table shows the rank-biserial correlations between the specialists' assessments and the ranks calculated via our method, alongside the MWU test results produced for two groups (the samples labelled as homogeneous and heterogeneous by each expert). Importantly, the experts labelled heterogeneous samples as 1 and homogeneous samples as 0. Therefore, a positive correlation means that as the attribute value increased, the corresponding sample tended to be more heterogeneous. The p-values indicate the likelihood of obtaining the observed results, or even more extreme outcomes, under the assumption that there was no difference between the groups. Significant differences between the groups were observed when the mean, standard deviation, and coefficient of variation were used. The only cases that did not achieve p-values lower than 0.05 were the comparisons between the mean and standard deviation attributes with the assessments of Expert 4. As expected, the results obtained via kurtosis were weaker, but a significant difference was still observed between the groups assessed by Experts 1 and 2, who evaluated the largest sample size.

Sample size discrepancies are important, and it should also be noted that the experts evaluated different sets of samples, with only partial overlap. This could have resulted in some experts encountering samples with less clear heterogeneity or homogeneity, potentially affecting the comparison outcomes of their assessments. Notably, the correlations with Expert 4 were weaker than those observed with the other experts. The results demonstrate that our method performed well when the mean, standard deviation, and coefficient of variation attributes Were utilized. The latter attribute produced statistically significant correlations with all the experts, effectively rejecting the null hypothesis of no correlation. This outcome is particularly noteworthy given that considerable divergences were observed among the experts themselves. The rank-biserial correlation coefficient and the MWU test were used instead of the point-biserial correlation coefficient and the t test because the assumptions of normality and equal variances were not satisfied in our data.

\begin{table}[h!]
\centering
\caption{Rank-biserial correlation coefficients (\(\rho_{rb}\)) and p-values derived from the MWU test conducted between the results of our method and the heterogeneity assessments of the four experts for different statistical attributes. The table includes results for the mean, standard deviation, coefficient of variation, and kurtosis attributes. The significance levels are denoted by asterisks: *$\text{p-value} < 0.05$, **$\text{p-value} < 0.01$, ***$\text{p-value} < 0.001$.}
\label{tab:rankbiserial}
\medskip

{\small
\setlength{\tabcolsep}{4pt}

\begin{tabular}{l
                l l
                l l
                l l
                l l}
\toprule
& \multicolumn{2}{c}{\textbf{Mean}} 
& \multicolumn{2}{c}{\textbf{Std. Dev.}} 
& \multicolumn{2}{c}{\textbf{Coeff. of Var.}} 
& \multicolumn{2}{c}{\textbf{Kurtosis}} \\
\cmidrule(lr){2-3} \cmidrule(lr){4-5} \cmidrule(lr){6-7} \cmidrule(lr){8-9}
& \(\rho_{rb}\) & p-value 
& \(\rho_{rb}\) & p-value 
& \(\rho_{rb}\) & p-value 
& \(\rho_{rb}\) & p-value \\
\midrule

\textbf{Expert 1} 
& 0.80 & $1.28 \times 10^{-13}$*** 
& 0.73 & $8.71 \times 10^{-12}$*** 
& 0.73 & $9.21 \times 10^{-12}$*** 
& 0.31 & $3.77 \times 10^{-3}$** \\

\textbf{Expert 2} 
& 0.51 & $1.63 \times 10^{-5}$*** 
& 0.71 & $1.15 \times 10^{-9}$*** 
& 0.73 & $4.69 \times 10^{-9}$*** 
& 0.26 & $2.73 \times 10^{-2}$* \\

\textbf{Expert 3} 
& 0.68 & $4.32 \times 10^{-3}$** 
& 0.75 & $1.64 \times 10^{-3}$** 
& 0.68 & $4.33 \times 10^{-3}$** 
& 0.29 & $2.38 \times 10^{-1}$ \\

\textbf{Expert 4} 
& 0.33 & $7.60 \times 10^{-2}$ 
& 0.36 & $5.70 \times 10^{-2}$ 
& 0.40 & $3.26 \times 10^{-2}$* 
& 0.16 & $4.00 \times 10^{-1}$ \\
\bottomrule
\end{tabular}
}
\end{table}

Table \ref{tab:expertglcm} presents the rank-biserial correlation coefficients between the GLCM attributes calculated for the micro-CT images and the same set of samples labelled as homogeneous or heterogeneous by each expert. Since the GLCM method is a 2D technique, the attributes of the 3D micro-CT images were obtained by averaging the results obtained across 100 equally spaced slices per view, using a distance of 1 and angles of 0 and 90 degrees. Energy consistently showed a negative correlation with our rank, indicating that samples that were more likely to be homogeneous had higher energy values. Similarly, homogeneity and dissimilarity showed correlations in the expected directions: homogeneity produced predominantly negative correlations, whereas dissimilarity produced positive correlations. This was in line with the attribute descriptions presented in Table \ref{tab:glcm_attributes}. However, despite these expected trends, the magnitudes of the correlations were relatively small, and none were statistically significant ($\text{p-value} < 0.05$). This means that we could not reject the hypothesis that both groups, labelled homogeneous and heterogeneous, shared the same distribution. Compared with the results obtained by our method, which revealed significant differences between the homogeneous and heterogeneous samples labelled by most experts across three different attributes, our method better captured the distinct behaviours of these groups, as indicated by the MWU test.

\begin{table}[h!]
\centering
\caption{Comparison among the rank-biserial correlation coefficients (\(\rho_{rb}\)) and p-values derived from MWU test between the GLCM attributes and the heterogeneity assessments provided by the four experts. The table includes results for the dissimilarity, energy, and homogeneity attributes.}
\label{tab:expertglcm}
\medskip

\begin{tabular}{l ll ll ll}
\toprule
& \multicolumn{2}{c}{\textbf{Dissimilarity}} & \multicolumn{2}{c}{\textbf{Energy}} & \multicolumn{2}{c}{\textbf{Homogeneity}} \\
\cmidrule(lr){2-3} \cmidrule(lr){4-5} \cmidrule(lr){6-7}
& \textbf{$\rho_{rb}$} & p-value & \textbf{$\rho_{rb}$} & p-value & \textbf{$\rho_{rb}$} & p-value \\

\midrule

\textbf{Expert 1} & 0.01 & $9.11 \cdot 10^{-1}$ & -0.09 & $4.31 \cdot 10^{-1}$ & 0.02 & $8.72 \cdot 10^{-1}$ \\
\textbf{Expert 2} & 0.07 & $5.55 \cdot 10^{-1}$ & -0.17 & $1.65 \cdot 10^{-1}$ & -0.06 & $6.31 \cdot 10^{-1}$ \\
\textbf{Expert 3} & 0.15 & $5.53 \cdot 10^{-1}$ & -0.23 & $3.96 \cdot 10^{-1}$ & -0.06 & $8.29 \cdot 10^{-1}$ \\
\textbf{Expert 4} & -0.12 & $5.26 \cdot 10^{-1}$ & -0.16 & $3.86 \cdot 10^{-1}$ & 0.06  & $7.66 \cdot 10^{-1}$ \\

\bottomrule
\end{tabular}

\end{table}

\section{Conclusion}\label{sec:conclusion}

Reservoir heterogeneity plays a crucial role in assessing the viability of exploration and production processes in oil and gas fields. Traditional methods, such as visual core inspection and laboratory measurements, are time-consuming, expensive, and prone to subjectivity. Moreover, segmentation-based approaches add another layer that can induce potential errors. In this study, we proposed an automated method for quantifying rock heterogeneity using only micro-CT images. The approach involves systematically dividing the input images into subvolumes and calculating statistical attributes within each subvolume. Entropy, a statistical measure of uncertainty and information content, is then computed for each attribute to assess the degree of heterogeneity, providing an objective textural heterogeneity measure.

The proposed method was applied to a dataset consisting of 4,935 micro-CT images of 4,744 cylindrical plug samples derived from Brazilian reservoirs. The results demonstrated that the textural heterogeneity measure is sensitive to variations arising from lithological changes, diagenetic minerals, porous structures, and potential artifacts. The entropy values calculated from the mean, standard deviation, and coefficient of variation attributes were found to be consistent with each other and strongly correlated with expert assessments of heterogeneity, as evidenced by the significant p-values obtained through statistical tests. In contrast, it was also shown that this outcome did not hold true for all attributes, highlighting the importance of conducting attribute selection. The performance of the proposed method was further validated by comparing its results to those obtained via GLCM attributes, where our method consistently produced stronger and more reliable correlations with the expert evaluations.

The automated method developed in this study provides an objective measure of rock textural heterogeneity via micro-CT images. Owing to its sole reliance on images, this method addresses the common limitations of the traditional heterogeneity assessment techniques found in the literature and offers a scalable solution for reservoir characterization. Future work should further evaluate the impacts of finer or coarser division schemes and attribute selection on the accuracy of heterogeneity assessments, with a focus on utilizing more complex attributes that may better capture the spatial distribution within each subvolume. Additionally, efforts should be made to address challenges associated with high-density materials and artifacts, which highly affect the heterogeneity measure. Implementing advanced image processing techniques to mitigate these effects could enhance the robustness of the method. Correlating the measure of heterogeneity with petrophysical properties could also provide deeper insights into the applicability of the proposed method in reservoir characterization.

\backmatter


\subsection*{Acknowledgements}

The authors would like to thank PETROBRAS for providing financial support for this project as well the data used in this work. We would also like to thank the Conselho Nacional de Desenvolvimento Científico e Tecnológico (CNPq) and Coordenação de Aperfeiçoamento de Pessoal de Nível Superior (CAPES) for providing scholarships to the authors. We also express our gratitude to the Laboratório Nacional de Computação Científica (LNCC) for providing the computational resources required for this research.

\subsection*{Availability of data and material} This dataset is available with permission from the Brazilian National Agency of Petroleum, Natural Gas and Biofuels\footnote{https://www.gov.br/anp/pt-br}. The algorithm's Python code can be accessed through the GitHub link \url{https://github.com/luancvieira/Textural-Heterogeneity-Measure}.

\section*{Declarations}

\subsection*{Conflicts of interest/competing interests} The authors declare that they have no competing interests.

\subsection*{Author Contributions}\textbf{Luan Coelho Vieira da Silva:} Conceptualization, Methodology, Software, Formal analysis, Writing – original draft. \textbf{Julio de Castro Vargas Fernandes:} Conceptualization, Methodology, Formal analysis, Writing – original draft. \textbf{Felipe Bevilaqua Foldes Guimarães:} Conceptualization, Software, Formal analysis. \textbf{Pedro Henrique Braga Lisboa:} Conceptualization, Software. \textbf{Carlos Eduardo Menezes dos Anjos:} Conceptualization, Methodology, Writing - Review \& Editing. \textbf{Thais Fernandes de Matos:} Conceptualization, Data curation, Formal analysis, Writing – original draft. \textbf{Marcelo Ramalho Albuquerque:} Conceptualization, Writing – Review \& Editing, Supervision. \textbf{Rodrigo Surmas:} Conceptualization, Data curation, Writing – Review \& Editing, Supervision. \textbf{Alexandre Gonçalves Evsukoff:} Conceptualization, Writing – Review \& Editing, Project administration.

\end{document}